\documentclass[aps, prd, superscriptaddress, nofootinbib, preprintnumbers]{revtex4}
\usepackage{amsmath}
\usepackage{graphicx}

\begin{document}
 \title{{\bf Gluonic Effects on $g-2$: Holographic View} 
 \vspace{5mm}}
 \author{Masafumi Kurachi} \thanks{\tt kurachi@kmi.nagoya-u.ac.jp}
      \affiliation{ Kobayashi-Maskawa Institute for the Origin of Particles and 
the Universe (KMI) \\ 
 Nagoya University, Nagoya 464-8602, Japan.}
\author{Shinya Matsuzaki}\thanks{\tt synya@hken.phys.nagoya-u.ac.jp}
      \affiliation{ Institute for Advanced Research, Nagoya University, Nagoya 464-8602, Japan.}
      \affiliation{ Department of Physics, Nagoya University, Nagoya 464-8602, Japan.}
\author{{Koichi Yamawaki}} \thanks{
      {\tt yamawaki@kmi.nagoya-u.ac.jp}}
      \affiliation{ Kobayashi-Maskawa Institute for the Origin of Particles and 
the Universe (KMI) \\ 
 Nagoya University, Nagoya 464-8602, Japan.}

\begin{abstract}
\vspace{5mm}
We study ``gluonic effects" (gluon condensation effects) on the hadronic leading 
order (HLO) contributions to the anomalous magnetic moment ($g-2$) of 
leptons, based on a holographic model having explicit gluonic mode introduced for consistency with the operator product expansion of QCD.  
We find gluonic enhancement of HLO contributions to the muon $g-2$ by 
about 6\%, which nicely fills in the gap between the holographic estimate without gluonic 
effects and the phenomenological one using 
the experimental data as inputs. 
Similar calculations including the gluonic effects for the electron and the 
tau lepton $g-2$ are also carried out in good agreement 
with the phenomenological estimates. 
We then apply our holographic estimate to the Walking Technicolor (WTC) 
where large techni-gluonic effects were shown to be vital 
for the Technidilaton, (pseudo) Nambu-Goldstone boson
of the (approximate) scale symmetry of WTC,  
to be naturally as light as 125 GeV.
 It is shown that the value of the techni-HLO contributions to the muon $g-2$ is 
10-100 times enhanced by inclusion of the same amount of the gluonic effects as that realizing the 
125 GeV Technidilaton, although  such an enhanced techni-HLO contribution is still negligibly small 
compared with the current deviation of the Standard Model prediction of the muon $g-2$ from the
experiments.  
The techni-HLO contributions to the tau lepton $g-2$ is 
also discussed, suggesting  
a possible phenomenological relevance to be tested by the future experiments.

\end{abstract} 
\maketitle

\section{Introduction} 
Holography, based on AdS/CFT (anti-de-Sitter space/conformal field theory) correspondence~\cite{Maldacena:1997re}, 
 has been extensively used to analyze strongly coupled gauge systems. 
For instance, QCD can be reformulated based on the gauge-gravity duality,  
either in the bottom-up approach~\cite{DaRold:2005zs,Erlich:2005qh} modeled as a five-dimensional 
gauge theory defined in an AdS background, 
or in the top-down approach~\cite{Sakai:2004cn} starting with a stringy setting. 
Such models of holographic QCD have succeeded well in reproducing 
several important properties for QCD hadrons within a theoretically expected size of 
uncertainties. 
In particular, the bottom-up approach can be made to reproduce the QCD in the all energy region, from the 
high energy behavior 
through the operator product expansion (OPE) down to 
the low energy resonance physics: The correct power behaviors in OPE for the chiral condensate and 
the gluon condensate are realized by introduction of the bulk (chiral non-singlet) scalar field, $\Phi_S$, corresponding to the $\bar q q$ operator and (chiral singlet) bulk gluonic field, $\Phi_G$,  respectively~\cite{Haba:2010hu}, 
which is contrasted to the top-down approach having high energy behavior completely different than that of QCD. 
The bottom-up holography so constructed can provide us with novel insights into 
the strong dynamics through 
the highly nontrivial effects of the gluonic dynamics: 
It reproduces nicely the known value of the gluon condensate which is otherwise zero, and the mass of $a_1$ meson of the right magnitude.

Such a bottom-up holographic QCD model having gluon condensation effects (hereafter, we will use the phrase ``gluonic effects" to refer to the gluon condensation effects) $\Phi_G$
was further applied~\cite{Haba:2010hu} to the Walking Technicolor (WTC) model~\cite{Yamawaki:1985zg} which
has an approximate scale symmetry and a large anomalous 
dimension $\gamma_m=1$, and further predicts a light composite Higgs dubbed ``Technidilaton (TD)'' as  a pseudo Nambu-Goldstone boson of the approximate scale symmetry.
The holographic WTC of Ref.~\cite{Haba:2010hu} was formulated through
a simple replacement of $\gamma_m=0$ for QCD by 
$\gamma_m=1$ in the mass parameter of the bulk scalar $\Phi_S$ in the holographic QCD with $\Phi_G$,  which was 
found~\cite{Matsuzaki:2012xx}  to have more intriguing gluonic effects than in those of the holographic QCD:
Large gluonic effects due to $\Phi_G$ in WTC actually realize an idealized limit where 
TD~\cite{Yamawaki:1985zg, Bando:1986bg}  
as a flavor-singlet scalar fermionic bound state (lowest Kaluza-Klein (KK) mode of $\Phi_S$)\footnote{
Note that a scalar techni-glueball as the lowest KK mode of $\Phi_G$ has no direct relevance to the electroweak symmetry breaking, 
since the technigluons carry no electroweak charge, and hence cannot be identified with the TD, or composite Higgs, whose constituents
must carry the electroweak charge.
} has a vanishingly small mass compared with the 
typical symmetry breaking scale $4\pi F_\pi$ ($F_\pi$: techni-pion decay constant), 
and hence it can naturally be as light as 125 GeV to be identified as the  Higgs
boson discovered at the LHC~\cite{Chatrchyan:2012ufa}. 
This is in sharp contrast to the earlier nonperturbative  
studies based on the ladder approximation~\cite{Harada:2003dc,Shuto:1989te} without 
nonperturbative gluonic dynamics,  which yields a substantially smaller scalar mass than in the QCD case, 
though it does not have such an idealized  massless limit and hence no natural 
framework for the light TD~\footnote{We actually have a formal limit of massless TD through the 
ladder estimate for the Partially Conserved Dilatation Current (PCDC) relation, 
which involves mass $M_\phi$ and the decay constant $F_\phi$ of the TD $\phi$
only in the form of a product, $M_\phi  F_\phi ={\cal O} (F_\pi^2)$,  (not separately). 
This implies that  a massless TD limit $M_\phi/F_\pi \rightarrow 0$ is formally realized 
only when $F_\phi/F_\pi \rightarrow \infty$, or the coupling vanishes (decoupled dilaton)~\cite{Haba:2010hu,Hashimoto:2010nw}.
Nevertheless, it so happened that  the mass estimated by the ladder PCDC  
actually can be parametrically tuned to be 125 GeV  in such a way as 
to be consistent with the current LHC data, being still far from the decoupling limit~\cite{Matsuzaki:2012gd}.
}.  
Therefore, proper inclusion of gluonic effects 
is important not only for the study of QCD, but also, or more significantly, for 
the realistic WTC calculations.

Holographic computations are generically done through evaluating Green 
functions constructed from (techni-)quark and gluon currents, so that the 
main outputs are made from the current correlators, including full information 
on masses and couplings for the associated mesons and glueball coupled 
to those currents. Once the vector current correlator is obtained from 
the holographic calculation, it is possible to translate it into the 
electromagnetic current correlator, $\Pi_{\rm em}(Q^2)$, from which we can estimate 
the (techni-)hadronic leading order (HLO) contribution to the 
anomalous magnetic moment of the muon (the muon $g-2$).  
 Actually, in Ref.~\cite{Hong:2009jv} a holographic estimation of such a QCD 
 HLO contribution was done 
 based on an earlier bottom-up holographic model~\cite{Erlich:2005qh}, 
in which, however, effects from the gluon condensation are not incorporated. 
 
In this paper, we study the gluonic effects on the muon $g-2$ in QCD and WTC 
through the (techni-)HLO contribution, based on a recently published holographic model 
in the bottom-up approach~\cite{Haba:2010hu,Matsuzaki:2012xx}.  
The model has 3 holographic parameters ($z_m,\, \xi,\, G$) to be explained later; $\xi$ 
and $G$, roughly corresponding to the chiral condensate and the gluon condensate, respectively, at the infrared brane located at $z_m$.
 By fixing the values $f_\pi=92.4$ MeV and 
$M_\rho =775.49$ MeV as inputs,  we have $\xi$ and $z_m$ as functions of $G$.
Accordingly, we 
show all other hadronic observables given as functions of a single holographic gluonic parameter $G$: 
While the mass of the flavor-singlet scalar 
$f_0(1370)$ and the scalar glueball are insensitive to the value of $G$, 
the mass of $a_1$ meson and the  gluon condensate which are fairly sensitive to $G$ so that we can determine 
the value of $G\simeq 0.25$ nicely fitting the reality of all the observables studied~\cite{Haba:2010hu,Matsuzaki:2012xx}.

We then estimate the HLO contributions  including the gluon-condensation effect, 
based on a formula  for the HLO directly evaluated by the current correlator in the 
(theoretically more tractable) {\it space-like momentum} in contrast to
the conventional one  converting it to the 
dispersion integral from the time-like contributions where the experimental data are available. 
The results show that the proper inclusion 
of gluon-condensation effect $G\simeq 0.25$ causes about 6\% enhancement of the HLO 
contribution to the muon $g-2$:
$a_\mu^{\rm HLO}|_{N_f=2} \simeq  
505 \times 10^{-10}$, 
$a_\mu^{\rm HLO}|_{N_f=3} \simeq  
606 \times 10^{-10}$, which are compared to the 
previous calculation~\cite{Hong:2009jv}  without 
gluon-condensation effect $G=0$: $a_\mu^{\rm HLO}|_{N_f=2} \simeq 
 476 \times 10^{-10}$,  
$a_\mu^{\rm HLO}|_{N_f=3} \simeq 
 571 \times 10^{-10}$, respectively.  
 This enhancement makes the prediction of 
holographic calculation of the HLO contribution very close to the known 
value~\cite{Hagiwara:2011af} from the phenomenological estimate using the experimental inputs
$a_\mu^{\rm HLO}|_{\pi^+\pi^-}=(504.2 \pm 3.0) \times 10^{-10}$ and 
$a_\mu^{\rm HLO}|_{\rm full}=(694.9 \pm 4.3)\times 10^{-10}$, respectively. 
We also calculate the HLO contributions to 
the electron and the tau lepton $g-2$, and show 
that the holographic predictions are 
quite consistent with the known values as well.

Considering the fact that the 
energy scales important for determining
the HLO effects are hierarchically different depending on the lepton species, 
it is remarkable that the holographic calculations, with proper 
inclusion of gluon-condensation effect, can reproduce 
HLO contributions to $g-2$ for all the leptons.
This means that with inclusion of the gluonic contributions the holographic calculation of $\Pi_{\rm em}(Q^2)$  is 
quite reliable in a {\it wide range of energy scale},  and remarkably in  
the (theoretically tractable) continuous space-like momentum, not just  in the range of the (discrete) time-like momentum where the resonance parameters are fitted in the conventional holographic
studies in the zero-width approximation (large $N_c$ limit).

Encouraged by this success, 
we apply the same holographic calculations of the HLO-type contributions to the $g-2$ from the  
WTC, with $\gamma_m=0$ in QCD simply replaced by $\gamma_m=1$ through the bulk scalar mass.
We first show the gluonic effects on various observables 
in WTC, including the mass of the flavor-singlet scalar meson $M_\phi$ identified with TD,
which can be as light as the 125 GeV boson discovered at the LHC for a large gluonic effect with 
$G\simeq 10$~\cite{Matsuzaki:2012xx}. 
Then we study the effect of such a large $G$ on the muon $g-2$ 
through the techni-HLO contribution, which we find as large as 100 times of the value for $G=0$:
Imposing  a typical constraint on the Peskin-Takeuchi $S$ parameter~\cite{Peskin:1990zt}  from the electroweak (EW) precision test, $S=0.1$, 
the enhancement is about 10 
times and by relaxing the constraint on $S$ parameter as $0.1<S<1.0$ for the reason described in the text, 
the techni-HLO contribution can be further enhanced by another factor of about 10. 
For all such enhancements, however,  we show that the techni-HLO contributions from WTC dynamics is 
negligibly small compared to 
 the HLO contributions from QCD. 
It is then very unlikely that 
the contributions from WTC dynamics can explain 
the inconsistency by about 3.3 $\sigma $ between the experimental value 
of the muon $g-2$ and the Standard Model (SM) prediction of it. 
We also mention the techni-HLO contribution 
to the tau $g-2$, which may become relevant in future experiments. 

The paper is organized as follows:
In the next section, we review the model \cite{Haba:2010hu} 
and formulas which are needed for the calculations of various physical 
quantities studied in this paper. 
In section~\ref{sec:QCD}, we study the gluonic effect of the muon $g-2$ through the holographic QCD
calculation of HLO contribution, based on the successful inclusion of the gluonc effects
on the various hadonic observables in Ref.~\cite{Haba:2010hu}. 
In section~\ref{sec:WTC}, we use similar 
holographic calculations for the study of WTC effects on the $g-2$.
Section~\ref{sec:conclusions} is devoted to discussions and 
summary of the paper.

\section{Holographic model and formulas} 
 
The model~\cite{Haba:2010hu} we shall employ is based on deformations of 
a bottom-up approach for successful holographic dual 
of QCD~\cite{DaRold:2005zs,Erlich:2005qh}.
The model is described as $SU(N_f)_L \times SU(N_f)_R$ gauge theory 
defined on the five-dimensional AdS space-time, which is characterized by 
the metric $ds^2= g_{MN} dx^M dx^N 
= \left(L/z \right)^2\big(\eta_{\mu\nu}dx^\mu dx^\nu-dz^2\big)$ 
with $\eta_{\mu\nu}={\rm diag}[1, -1, -1,-1]$. 
Here,  $M$ and $N$ ($\mu$ and $\nu$) represent five-dimensional (four-dimensional) 
Lorentz indices, and $L$ denotes the curvature radius of the AdS background.
The fifth direction, denoted as $z$, is compactified on an interval extended 
from the ultraviolet (UV) brane located at $z=\epsilon$ 
to the infrared (IR) brane at $z=z_m$, i.e., $ \epsilon \leq z \leq z_m  $. 
The UV cutoff $\epsilon$ will be taken to be 0 after all calculations are done.  
  In addition to the bulk left- ($L_M$) and right- ($R_M$) gauge fields, 
we introduce a bulk scalar $\Phi_S$ which transforms as a bifundamental 
representation field under the $SU(N_{f})_L \times SU(N_{f})_R$ gauge symmetry, 
and therefore 
it is considered to be dual to the quark bilinear operator $\bar{q} q$. 
The mass-parameter $m_{\Phi_S}$ is then related to $\gamma_m$ 
as 
\begin{equation}
m_{\Phi_S}^2=- \frac{(3-\gamma_m)(1+ \gamma_m)}{L^2}\,.
\end{equation}
We take  
$\gamma_m = 0$ when we apply this holographic model to the 
study of the actual QCD, while we take $\gamma_m = 1$ when we consider 
the model as a dual of WTC.  
The action of the model is given as ~\cite{Haba:2010hu} 
\begin{equation} 
  S_5 = S_{\rm bulk} + S_{\rm UV} + S_{\rm IR} 
  \,, \label{S5}
\end{equation}
where $S_{\rm bulk}$ denotes the five-dimensional bulk action, 
\begin{eqnarray} 
  S_{\rm bulk} 
  &=& 
  \int d^4 x \int_\epsilon^{z_m} dz 
  \sqrt{g} 
  \frac{1}{g_5^2} \, e^{c_G g_5^2 \Phi_G} 
 \Bigg[ 
\frac{1}{2} \partial_M \Phi_G \partial^M \Phi_G 
\nonumber \\ 
&& 
+ {\rm Tr}[D_M \Phi_S^\dag D^M \Phi_S - m_{\Phi_S}^2 \Phi_S^\dag \Phi_S ] 
\nonumber \\ 
&&
  - \frac{1}{4} {\rm Tr}[L_{MN}L^{MN} + R_{MN}R^{MN}] 
 \Bigg] 
 \,, \label{S:bulk}
\end{eqnarray}
and $S_{\rm UV, IR}$ the boundary actions which are given in Ref.~\cite{Matsuzaki:2012xx}.   
The covariant derivative acting on $\Phi_S$ in Eq.(\ref{S:bulk}) 
 is defined as $D_M\Phi_S=\partial_M \Phi_S+iL_M\Phi_S-i\Phi_S R_M$, where 
$L_M(R_M)\equiv L_M^a(R_M^a) T^a$ with $T^a$ being the generators of 
$SU(N_{f})$ which are normalized as 
${\rm Tr}[T^a T^b]=\delta^{ab}$. 
$L(R)_{MN}$ is the five-dimensional field strength which is defined as 
$L(R)_{MN} = \partial_M L(R)_N - \partial_N L(R)_M 
 - i [ L(R)_M, L(R)_N ]$, and 
$g$ is defined as $g={\rm det}[g_{MN}]= (L/z)^{10}$. 
A salient feature of the model is the extra bulk scalar $\Phi_G$  
introduced in Eq.(\ref{S:bulk}) in order to reproduce the correct asymptotic behavior of the
QCD. This is a bulk field which is dual to the gluon condensate $\langle \alpha_s G_{\mu\nu}^2\rangle$ in QCD. 
Here, $\alpha_s$ is related to the QCD gauge couping $g_s$ by 
$\alpha_s = g^2_s/(4\pi)$. 
Since $\langle \alpha_s G_{\mu\nu}^2 \rangle$ is a singlet under 
the chiral $SU(N_{f})_L \times SU(N_{f})_R$ symmetry, 
the dual-bulk scalar $\Phi_G$ has to be a real field. 
We take ${\rm dim}(\alpha_s G_{\mu\nu}^2)=4$, thus  
the corresponding bulk-mass parameter becomes $m_{\Phi_G}^2=0$.
The gauge coupling $g_5$ and a parameter $c_G$ appearing in the action 
are fixed as  
\begin{equation}
\frac{L}{g_5^2} = \frac{N_{c}}{12\pi^2},\quad c_G = -\frac{L}{16\pi g_5^2} = - \frac{N_c}{192 \pi^3}\,,
\label{matching}
\end{equation}
so that the model reproduces the OPE in QCD (see later discussions)~\cite{Haba:2010hu}.

We shall begin with the bulk scalar sector in Eq.(\ref{S:bulk}). 
The bulk scalar fields $\Phi_S$ and $\Phi_G$  
are parametrized as   
\begin{equation}
\Phi_S(x,z) 
= \frac{1}{\sqrt{2}} \left(v_S(z)+ \frac{\sigma_S(x,z)}{\sqrt{N_f}}\right) e^{2i \pi(x,z)/v_S(z)}, 
\end{equation}
\begin{equation} 
\chi_G(x,z) \equiv e^{c_G g_5^2\Phi_G/2 } =v_{\chi_G}(z)\, e^{\sigma_{\chi_G}(x,z)/v_{\chi_G}(z)},
\end{equation}  
with the vacuum expectation values (VEVs), $v_S=\sqrt{2} \langle \Phi_S \rangle$ and 
$v_{\chi_G} = \langle \chi_G \rangle$. 
We hereafter disregard (techni-)pion fields $\pi$ which will not be relevant for the present study. 
The boundary conditions for $v_S(z)$ are~\cite{Haba:2008nz,Haba:2010hu,Matsuzaki:2012xx}:  
\begin{eqnarray}  
v_S(\epsilon)   
&=& 
\Bigg\{ 
\begin{array}{cc}
\left(\frac{\epsilon}{L} \right) \,  c_S^{\gamma_m=0} M 
& \quad \textrm{for} \quad \gamma_m =0 \\ 
\left(\frac{\epsilon}{L} \right)^2 \log \frac{z_m^2}{\epsilon^2} \,  c_S^{\gamma_m=1} M 
& \quad \textrm{for} \quad \gamma_m=1
\end{array}
\,, \label{BC:vs} \\   
v_S(z_m) &=& \frac{\xi}{L}
\,,  
\end{eqnarray}  
where $M$ stands for the current mass of (techni-)quarks, and 
the IR value $\xi$ is related to the (techni-)quark condensate $\langle \bar{q}q \rangle$.  
The parameter $c_S^{\gamma_m}$ has been introduced which can arise 
from the ambiguity of the definition for the current mass $M$, and 
is fixed to be $c_S^{\gamma_m=0} = \sqrt{3}$ for QCD 
and $c_S^{\gamma_m=1} =\sqrt{3}/2$ for WTC, 
by matching the UV asymptotic form of the scalar current correlator 
to the form predicted from the operator product expansion~\cite{Matsuzaki:2012xx}.

The boundary conditions for $v_{\chi_G}(z)$ are taken for both QCD and WTC cases 
as    
\begin{eqnarray}  
v_{\chi_G}(\epsilon)
&=& 
e^{\frac{c_G}{2} \frac{g_5^2}{L} M'} = e^{-\frac{1}{32\pi} L M'}
\,, \label{BC:vchi}\\  
v_{\chi_G}(z_m)  &=& 1 + G
\,, 
\end{eqnarray}  
where $M'$ becomes an external source for the (techni-)gluon 
condensate operator $(\alpha_s G^2_{\mu\nu})$, 
and $G$ is a parameter which is associated with the (techni-)gluon condensate 
$ \langle \alpha_s  G_{\mu\nu}^2 \rangle$. 
The  solutions of the VEVs $v_S(z)$ and $v_{\chi_G}(z)$ in the limit 
where $M \to 0$ and $M' \to 0$ are given as~\cite{Haba:2010hu}  
\begin{eqnarray}  
v_S(z) &=& 
\Bigg\{ 
\begin{array}{cc} 
\frac{\xi(1+G)}{L} \frac{(z/z_m)^3}{1+G(z/z_m)^4} 
& \quad \textrm{for} \quad \gamma_m =0  \\ 
\frac{\xi(1+G)}{L} \frac{(z/z_m)^2}{1+G(z/z_m)^4} 
\frac{\log(z/\epsilon)}{\log(z_m/\epsilon)} 
& \quad \textrm{for} \quad \gamma_m =1
\end{array}
 \\ 
v_{\chi_G}(z) &=& 
1 + G \left( \frac{z}{z_m} \right)^4 
\,.  
\end{eqnarray}

One can solve the equations of motion for the bulk scalars $\sigma_S$ and $\sigma_{\chi_G}$ 
with the UV boundary conditions similar to those in Eqs.~(\ref{BC:vs}) and (\ref{BC:vchi}) 
with $M$ and $M'$ replaced by 
sources $(s(x), g(x))$ for the scalar and gluonic currents $(J_S(x), J_G(x))$. 
Putting their solutions back into the action $S_5$ in Eq.(\ref{S5}), 
one can then obtain the generating functional $W[s(x),g(x)]$ holographically dual to QCD or WTC.  
The chiral  condensate and gluon condensate ($\langle \bar{q}q \rangle$  and $\langle \alpha_s G^2_{\mu\nu} \rangle$) 
are thus calculated by performing $\delta W/\delta s(x)$ and $\delta W/\delta g(x)$, respectively: 
\begin{eqnarray} 
\langle \bar{q}q \rangle_{1/z_m} 
&=& 
- \frac{c_S^{\gamma_m} (3-\gamma_m) N_c}{12 \pi^2} \frac{\xi (1+G)}{z_m^3}
\,, \qquad 
c_S^{\gamma_m}
= \Bigg\{
\begin{array}{cc} 
\sqrt{3} & \qquad {\rm for} \qquad \gamma_m=0 \\ 
\sqrt{3}/2 & \qquad {\rm for} \qquad \gamma_m=1 
\end{array}
\,, \label{Gcond}\\  
\langle \alpha_s G^2_{\mu\nu} \rangle 
&=& \frac{32 N_c}{3\pi} \frac{G}{z_m^4} \,, 
\label{condensates}
\end{eqnarray} 
where $\langle \bar{q}q \rangle_{1/z_m}$ is the chiral condensate renormalized at $\mu=1/z_m$~\cite{Haba:2008nz}. 
The chiral condensate 
renormalized at generic scale $\mu$ is given by $\langle \bar{q}q \rangle_{\mu}=Z_m^{-1}(\mu z_m)\cdot 
 \langle \bar{q}q \rangle_{1/z_m}$ with $Z_m^{-1}(\mu z_m)=(\mu z_m)^{\gamma_m}$. 
One can also calculate the scalar meson (TD) mass $M_S$ and glueball mass $M_G$ 
by evaluating the lowest poles of the scalar and gluonic current correlators~\cite{Matsuzaki:2012xx} 
to find the following eigenvalue equations: 
\begin{eqnarray} 
M_S &:&
\frac{3}{2} \xi^2 J_{1-\gamma_m}(M_S z_m) = (M_S z_m) J_{2-\gamma_m} (M_S z_m),
\label{MS}  \\  
 M_{G} &:& 
J_1(M_G z_m) = 0
\,, 
\label{MG}  
\end{eqnarray}
where $J$ is the Bessel function of the first kind.

We shall move on to the gauge sector in Eq.(\ref{S:bulk}).  
We define the five-dimensional vector and axial-vector gauge fields $V_M$ and $A_M$ 
as 
\begin{equation} 
 V_M = \frac{L_M + R_M}{\sqrt{2}} 
 \,, \qquad 
A_M = \frac{L_M-R_M}{\sqrt{2}}
\,. 
\end{equation} 
It is convenient to work with the gauge-fixing $V_z=A_z\equiv 0$ and take the boundary conditions 
$V_\mu(x,\epsilon)=v_\mu(x)$, $A_\mu(x,\epsilon)=a_\mu(x)$ and  $\partial_z V_\mu(x,z)|_{z=z_m}=\partial_z A_\mu(x,z)|_{z=z_m}= 0$, 
where $v_\mu(x)$  and $a_\mu(x)$ correspond to sources for the vector and axial-vector currents, respectively. 
We then solve the equations of motion for (the transversely polarized components of) $V_\mu(x,z)$ and $A_\mu(x,z)$ 
and substitute the solutions back into the action in Eq.(\ref{S:bulk}), to obtain the 
generating functional $W[v_\mu, a_\mu]$ holographically dual to QCD or WTC. 
Then the vector and the axial-vector current correlators are obtained
in a way similar to the case of the scalar sector.  
The correlators are defined as
\begin{eqnarray}
i \int d^4x e^{iqx} \langle 0 \vert\, {\rm T}\,  J_V^{a \mu}(x)\, J_V^{b \nu}(0)\, \vert 0 \rangle
 &=& \delta^{ab} \left( \frac{q^\mu q^\nu}{q^2} - \eta^{\mu\nu} \right) \Pi_V(-q^2),\\
i \int d^4x e^{iqx} \langle 0 \vert\, {\rm T}\,  J_A^{a \mu}(x)\, J_A^{b \nu}(0)\, \vert 0 \rangle
 &=& \delta^{ab} \left( \frac{q^\mu q^\nu}{q^2} - \eta^{\mu\nu} \right) \Pi_A(-q^2),
\end{eqnarray}
where the currents are defined as 
\begin{eqnarray}
J_V^{a\mu} &=& \bar{\psi} \left(\frac{T^a}{\sqrt{2}}\right) \gamma^\mu \psi,\\
J_A^{a\mu} &=& \bar{\psi} \left(\frac{T^a}{\sqrt{2}}\right) \gamma^\mu \gamma_5 \psi.
\end{eqnarray}
$\Pi_{V}(Q^2)$ and $\Pi_{A}(Q^2)$ (where $Q\equiv \sqrt{-q^2}$ is the Euclidean momentum)  
are expressed as 
\begin{eqnarray}
  \Pi_V(Q^2) &=& \frac{N_c}{12\pi^2} \frac{\partial_z V(Q^2, z)}{z} \Bigg|_{z=\epsilon \to 0}
  \,, \label{PiV}\\
    \Pi_A(Q^2) &=& \frac{N_c}{12\pi^2} \frac{\partial_z A(Q^2, z)}{z} \Bigg|_{z=\epsilon \to 0}
  \,,
\end{eqnarray} 
where the vector and axial-vector profile functions $V(Q^2,z)$ and $A(Q^2,z)$ are defined as $V_\mu(q,z)=v_\mu(q) V(q^2)$ and $A_\mu(q,z)=a_\mu(q) A(q^2)$ with the Fourier transforms of $v_\mu(x)$ and $a_\mu(x)$. 
These profile functions satisfy the following equations:   
\begin{equation} 
\left[ 
 - Q^2 + \omega^{-1}(z) \partial_z \omega(z) \partial_z  
\right] V(Q^2, z) = 0 ,
\label{EOM:PiV} 
\end{equation}
\begin{equation} 
\left[ 
 - Q^2 + \omega^{-1}(z) \partial_z \omega(z) \partial_z  - 2 \left( \frac{L}{z} \right)^2 [v_S(z)]^2
\right]A(Q^2, z) = 0 ,
\label{EOM:PiA}
\end{equation}
\begin{equation} 
\omega(z) \equiv \frac{L}{z} \left( 1 + G \left( \frac{z}{z_m} \right)^4 \right)^2 
\,,
\end{equation} 
with the boundary conditions 
$V(Q^2,z)|_{z=\epsilon \rightarrow 0}=A(Q^2,z)|_{z=\epsilon \rightarrow 0}=1$ 
and 
$\partial_z V(Q^2, z)|_{z=z_m}=\partial_z A(Q^2, z)|_{z=z_m}=0$.
It is worth mentioning that both the vector and the axial-vector 
current correlators involve gluonic effects through $G$. Most notably, we have  
\begin{equation}
\Pi_{V
}(Q^2) \Bigg|_{(1/z_m)^2 \ll Q^2 < (1/\epsilon)^2} 
= Q^2 \left[~\frac{L}{2 g_5^2}\log{Q^2} 
+ c_G \frac{2}{3} \frac{g_5^2}{L} \frac{\langle \alpha_s G_{\mu\nu}^2 \rangle}{Q^4} 
+\cdots
~\right]
\,, 
\label{PiV:complete0}
\end{equation}
in accord with the OPE in QCD:
\begin{eqnarray}
  \Pi_V^{(\rm QCD)} \left( Q^2 \right)\Bigg|_{\rm OPE} & = & 
Q^2 \left[\frac{N_c}{24 \pi^2}
\log \left( \frac{Q^2}{\mu^2} \right) 
 - \frac{1}{24 \pi}  \frac{  \langle \alpha_s G_{\mu\nu}^2 \rangle}{Q^4} 
+\cdots
\right] 
\,, 
\label{OPEVA}
\end{eqnarray} 
which was the basis for the parameter matching in Eq.(\ref{matching}). Were it not for the $\Phi_G$ term, we would not reproduce the correct gluon condensate term $1/Q^4$ in the QCD asymptotics
and also in WTC.  
Therefore the physical quantities 
which are related to these correlators, including the muon $g-2$, are influenced by the existence 
of gluonic dynamics.

The vector and axial-vector current correlators, $\Pi_V$ and $\Pi_A$, 
can be expanded in terms of towers of 
the vector and axial-vector resonances. 
We then identify the lowest poles for $\Pi_{V,A}$ as the (techni-)$\rho$ and 
$a_1$ mesons. Their masses, $M_\rho$ and $M_{a_1}$, 
 are calculated by solving the eigenvalue equations for 
the vector and axial-vector profile functions~\cite{Haba:2010hu}:  
\begin{equation} 
\left[ 
 M_\rho^2 + \omega^{-1}(z) \partial_z \omega(z) \partial_z  
\right] V(z) = 0 ,
\label{mrho:eq}
\end{equation}
\begin{equation} 
\left[ 
 M_{a_1}^2 + \omega^{-1}(z) \partial_z \omega(z) \partial_z  - 2 \left( \frac{L}{z} \right)^2 [v_S(z)]^2
\right]A(z) = 0 ,
\label{ma1:eq}
\end{equation}
with the same boundary conditions $V(\epsilon)=A(\epsilon)=0$ and $\partial_z V(z)|_{z=z_m}=\partial_z A(z)|_{z=z_m}=0$. 
We thus find $M_\rho$ and $M_{a_1}$ as functions of the model parameters $\xi$, $G$ and $\gamma_m$
with the overall scale set by $z_m$: 
$  
M_\rho =
z_m^{-1} \cdot \widetilde{M}_\rho(G) 
$ and  
$M_{a_1} =
  z_m^{-1} \cdot \widetilde{M}_{a_1} (\xi, G, \gamma_m) 
$.

The (techni-)pion decay constant is expressed as 
$
  f_\pi^2 = \Pi_V(0) - \Pi_A(0)  
$, 
which is, in the case of WTC, related to the EW scale $v_{\rm EW}\simeq 246$ GeV 
as $F_\pi = v_{\rm EW}/\sqrt{N_D}$.
Here, $N_D$ denotes the number of EW doublets, which is fixed to be 1 in 
the case of QCD.
  The present model enables us to express $f_\pi$ as a function of $\xi$, $G$, $\gamma_m$ and $z_m$~\cite{Haba:2010hu}: 
\begin{eqnarray} 
  f_\pi^2 &=&  
  \frac{N_{c}}{12\pi^2} \frac{\widetilde{F}^2(\xi, G,\gamma_m)}{z_m^2} 
 \,, \label{Fpi}
\end{eqnarray} 
where $\widetilde{F}^2 = - \partial_t A(Q^2=0,t=z/z_m)/t|_{t=\epsilon/z_m \to 0}$.

The $S$ parameter~\cite{Peskin:1990zt} is calculated from $\Pi_V$ and $\Pi_A$ as 
$ 
S= -16\pi L_{10}=-4\pi \, N_D \left[\Pi^\prime_V(0)- \Pi^\prime_A(0) \right]
$,  
where $\Pi^\prime_{V,A}(0) \equiv d\Pi_{V,A}(Q^2)/dQ^2 |_{Q^2=0}$. 
Thus $S$ is expressed as a function of two parameters  
 $\xi$ and $G$~\cite{Haba:2010hu} once $\gamma_m$ is fixed:  
\begin{eqnarray} 
 S =-16\pi L_{10}= \frac{N_D N_c}{3\pi} \int_{t_\epsilon}^1 \frac{dt}{t} v_\chi^2(t) \left( 1 - [A(Q^2=0, t=z/z_m)]^2 \right)  
 \,, \label{Spara} 
\end{eqnarray} 
 where $t_\epsilon = \epsilon/z_m (\to 0)$.

\begin{figure}[h] 
\begin{center} 
\includegraphics[width=6cm]{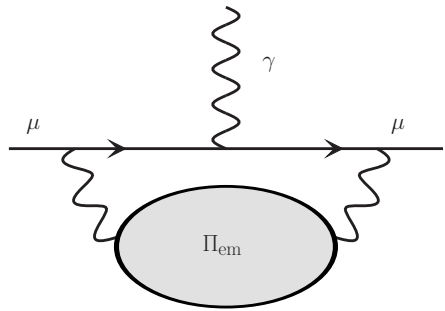} 
\end{center}
\caption{ 
An illustration of (techni-)HLO contributions to muon $g-2$. }  
\label{HLOfig}
\end{figure}

Once $\Pi_V(Q^2)$ is calculated in the holographic model, the electromagnetic current 
correlator $\Pi_{\rm em}(Q^2)$ is obtained as
\begin{equation}
 \Pi_{\rm em}(Q^2) = 2 \left( \sum_f q_f^2 \right)  \frac{\Pi_V(Q^2)}{Q^2},
\label{eq:Piem}
\end{equation}
where the summation runs over all the fermion flavors $f$ with their electromagnetic 
charge denoted as $q_f$. Here, we defined electromagnetic current correlator as 
\begin{equation}
i \int d^4x e^{iqx} \langle 0 \vert\, {\rm T}\,  J_{\rm em}^{\mu}(x)\, J_{\rm em}^{\nu}(0)\, \vert 0 \rangle
= \left( q^2 \eta^{\mu\nu} - q^\mu q^\nu \right) \Pi_{\rm em}(-q^2),
\end{equation}
where the electromagnetic current is defined as $
J_{\rm em}^\mu = \sum_f  q_f \bar\psi_f \gamma^\mu \psi_f$. 
With the electromagnetic current correlator given, 
the (techni-)hadronic leading order (HLO) contribution to the muon $g-2$ is  
calculated to be~\cite{Blum:2002ii} (For a graphical expression, see Fig.~\ref{HLOfig}.)

\begin{equation} 
\left.\left(
\frac{g-2}{2}
\right)\right\vert_{{\rm (techni)HLO}} 
\equiv
 a_\mu^{\rm (techni)HLO} =  4\pi^2 \left( \frac{\alpha_{\rm em}}{\pi} \right)^2 
\int_0^\infty d Q^2  f(Q^2) \Pi_{\rm em}^R(Q^2)
\,,
\label{eq:weight-1}
\end{equation}
where $\alpha_{\rm em}=e^2/(4\pi)$ with $e$ being the electromagnetic coupling constant. 
$\Pi_{\rm em}^R(Q^2)$ is the renormalized electromagnetic current correlator defined as 
$\Pi_{\rm em}^R(Q^2)\equiv \Pi_{\rm em}(Q^2)-\Pi_{\rm em}(0)$, and $f(Q^2)$ is a weight 
function which has the following form:
\begin{eqnarray}
f(Q^2) 
&=& 
\frac{m_\mu^2 Q^2 Z^3 (1-Q^2 Z)}{1 + m_\mu^2 Q^2 Z^2} 
\,, \nonumber \\ 
Z &\equiv&  \frac{\sqrt{Q^4 + 4 m_\mu^2 Q^2} - Q^2 }{2 m_\mu^2 Q^2}.
\label{eq:weight}
\end{eqnarray}
The formula can be rewritten in terms of $\Pi_V$ as follows:
\begin{equation} 
 a_\mu^{\rm (techni-)HLO} =  4\pi^2 \left( \frac{\alpha_{\rm em}}{\pi} \right)^2 
 {\cal N}
\int_0^\infty \frac{d Q^2}{Q^2}  f(Q^2) \left(\Pi_V(Q^2) - \Pi_V(0) \right), 
\label{amu:Hformula}
\end{equation}
where ${\cal N}$ is the prefactor in Eq.~(\ref{eq:Piem}), which takes the following 
values in the case of QCD and one-family technicolor model~\cite{Farhi:1979zx}:
\begin{eqnarray} 
{\cal N} \equiv 2 \sum_f q_f^2
= 
\left\{
\begin{array}{l}
10/9 \ \ \ {\rm QCD}\ (N_f=2)\\ 
\ 4/3 \ \ \ \, {\rm QCD}\ (N_f=3)\\
16/3\ \ \ \textrm{one-family technicolor}
\end{array} 
\right.
\end{eqnarray}

Now, all the physical quantities we are interested in have been expressed in terms of 
calculable holographic quantities. 
In the following two sections, for the case of QCD and WTC respectively, we evaluate
those physical quantities as functions of the model parameters $\xi, z_m, G$ and $\gamma_m$, especially focusing on 
their dependence on $G$ closely related to the gluon condensate.

\section{Gluonic effects in QCD}
\label{sec:QCD}

In this section, we apply the holographic calculations explained in the previous section 
for the estimation of various physical quantities in the real-life QCD. For this purpose, 
we take $\gamma_m=0$ so that the model correctly reproduces the UV asymptotic 
behavior of the QCD. After fixing the value of $\gamma_m$,
all the physical quantities are expressed as functions of 
$z_m$, $\xi$ and $G$. We use $M_\rho = 775.49$ MeV and $f_\pi=92.4$ MeV as inputs 
to fix two of these remaining model parameters. Then, from Eqs.~(\ref{mrho:eq}) and (\ref{Fpi}), 
one can express $z_m$ and $\xi$ as functions of $G$. Now, the remaining physical quantities, 
namely 
$M_{a_1}$, 
$\langle \frac{\alpha_s}{\pi} G_{\mu\nu}^2 \rangle$, 
$(-\langle \bar{q}q \rangle)^{1/3}$, 
$M_S$, 
$M_G$  
and 
$S=-16\pi L_{10}$ can be expressed as functions of a single 
parameter $G$ through
Eqs.~(\ref{ma1:eq}), (\ref{Gcond}), (\ref{condensates}), 
(\ref{MS}), (\ref{MG}) and (\ref{Spara}), which are depicted in 
Fig.~\ref{FpiMrhoFixed} by varying $G$ from 0 to 0.4: 
\begin{figure}[h]
\begin{center}
 \includegraphics[width=5.4cm]{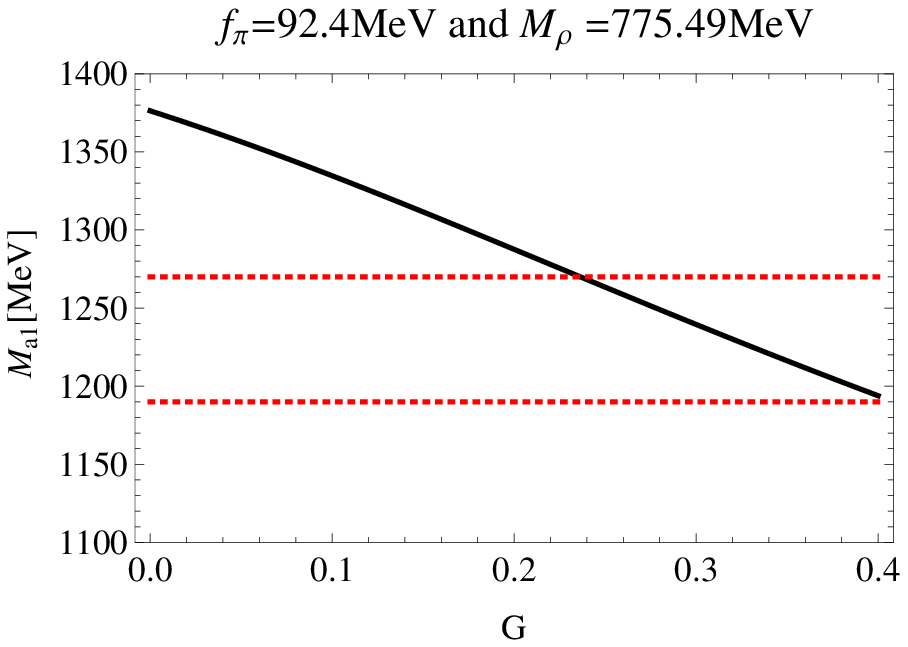}\ \ \ \ \ \ \ 
   \includegraphics[width=5.4cm]{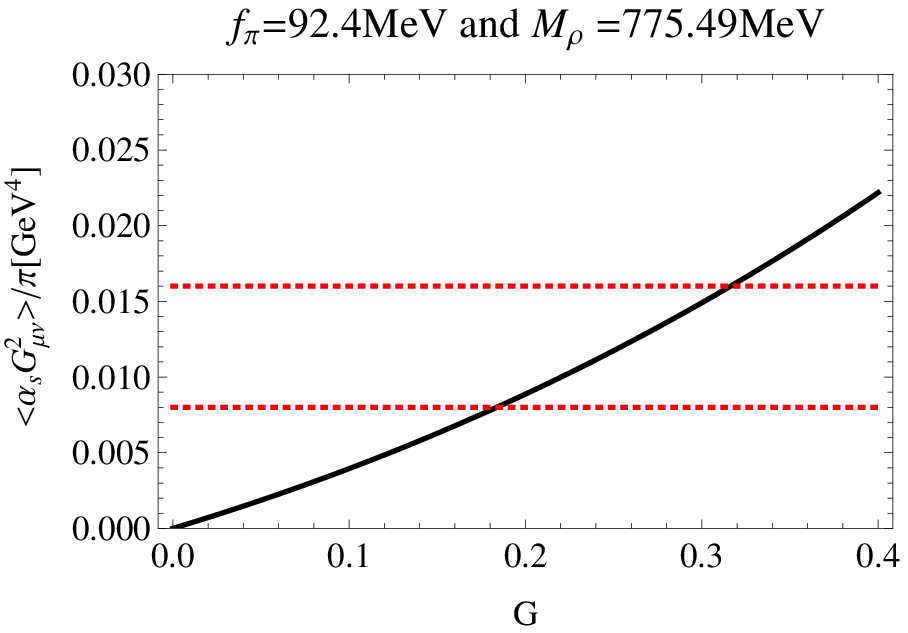} \ \ \ \ \ \ \ 
    \includegraphics[width=5.4cm]{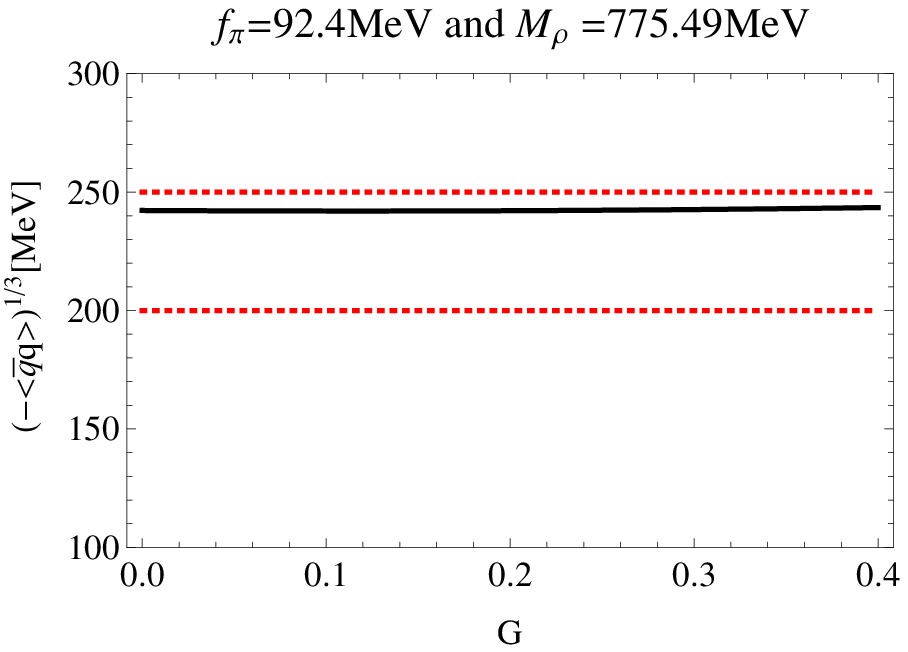}\vspace{5mm}\\
   \includegraphics[width=5.4cm]{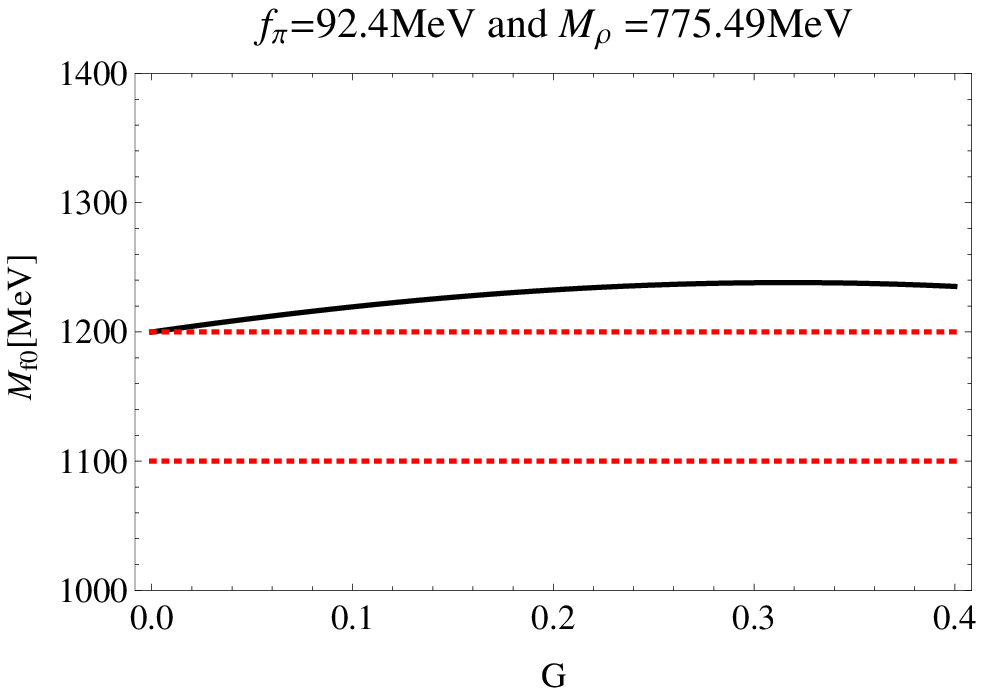} \ \ \ \ \ \ \ 
 \includegraphics[width=5.4cm]{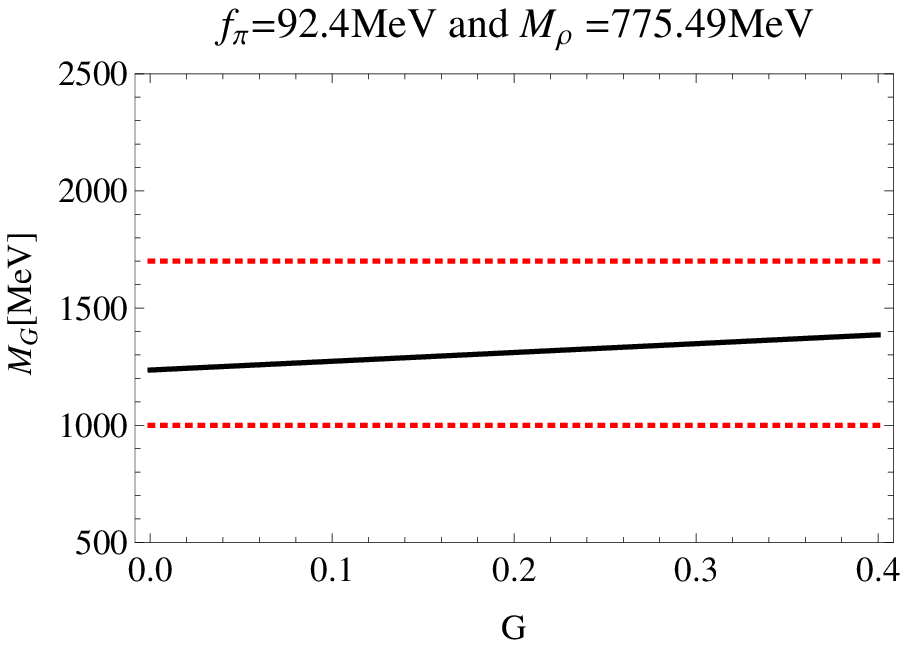}\ \ \ \ \ \ \ 
   \includegraphics[width=5.4cm]{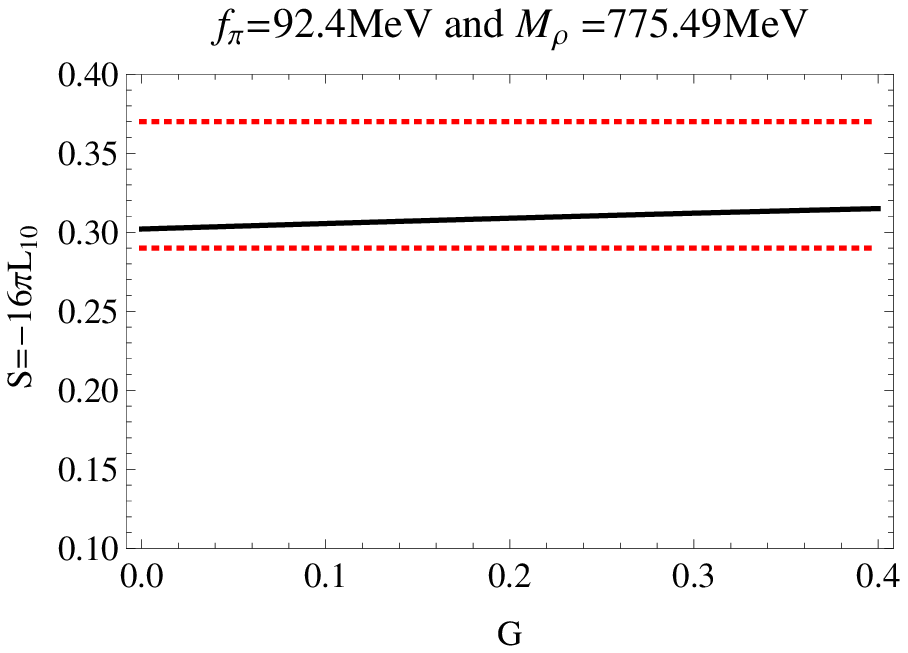}
\caption{ 
Plots of various QCD quantities as functions of $G$ with $f_\pi=92.4$ MeV and $M_\rho=775.49$ MeV fixed. 
Top-left: $M_{a_1}$~\cite{Beringer:1900zz}; 
Top-center 
: $\langle \frac{\alpha_s}{\pi} G_{\mu\nu}^2 \rangle$~\cite{Shifman:1978bx}; 
Top-right: $(-\langle \bar{q}q\rangle)^{1/3}$~\cite{Gasser:1982ap}; 
Bottom-left: $M_{S=f_0(1370)}$~\cite{Haba:2010hu};  
Bottom-center 
: $M_{G}$~\cite{Ochs:2013gi}; 
Bottom-right: $S=-16\pi L_{10}$~\cite{Harada:1992np}.
Two dashed red lines in  each plot 
correspond to observed upper and lower 
values of 1-$\sigma$ error band quoted in the corresponding references.
\label{FpiMrhoFixed}}
\end{center} 
 \end{figure} 
Observed values of those quantities are also indicated in the 
plots: two dashed red lines correspond to upper and lower values of 1-$\sigma$ error band.
 The ``observed" glueball mass $M_G$ has been taken from 
an expected mass range in lattice simulations~\cite{Ochs:2013gi}.  
As for the flavor-singlet two-quark bound state $S 
$ meson, 
we have chosen $f_0(1370)$ and estimated the ``observed" mass 
neglecting mixing with four-quark bound state $f_0(980)$.
(For more detailed discussions, see Ref.~\cite{Haba:2010hu}.)

The figures tell us that  
the $a_1$ meson mass $M_{a_1}$ and gluon condensate $\langle \frac{\alpha_s}{\pi} G_{\mu\nu}^2 \rangle$ 
highly depend on the change of $G$, 
while other quantities are rather insensitive to it, keeping values at around observed values of 
each quantity.  
From these results, we see the optimal value of  $G$ (and hence $z_m$ and $\xi$):~\cite{Haba:2010hu}  
\begin{equation} 
G \simeq 0.25 
\,,  \quad z_m^{-1} \simeq 347 \,{\rm MeV}\,, \quad \xi \simeq 3.1 \,\,,
\label{G:best} 
\end{equation} 
for the holographic model with $\gamma_m=0$ to reproduce 
the QCD observables. 
In  Table~\ref{fpi:mrho:ob:fixed:tab}
we show the results of holographic calculations for $G=0.25$ (and $G=0$ for comparison)  along with observed value 
of each quantity.

\begin{table} [h]
\begin{tabular}{|c||c|c|c|c|c|c|}
\hline 
 &  $M_{a_1}$ [MeV] 
&  $\langle \frac{\alpha_s}{\pi} G_{\mu\nu}^2 \rangle$ $[{\rm GeV}^4]$ 
&  $(- \langle \bar{q}q \rangle)^{1/3}$ [MeV] 
  &  $M_{S
  =f_0(1370)}$ [GeV]  
&  $M_{G}$ [GeV]  
&  $S= -16 \pi L_{10}$ \\ 
\hline \hline 
model ($G = 0$) 
& 1376
& 0 
& 277
& 1.20 
& 1.24 
& 0.30  
 \\
\hline 
model ($G = 0.25$) 
& 1264 
& 0.012 
& 277 
& 1.23 
& 1.33 
& 0.31  
 \\
\hline
measured 
& 1230 $\pm$ 40~\cite{Beringer:1900zz} 
& $0.012 \pm 0.004$~\cite{Shifman:1978bx}  
& $225 \pm 25$~\cite{Gasser:1982ap} 
& 1.1 -- 1.2~\cite{Haba:2010hu} 
& 1.0 -- 1.7~\cite{Ochs:2013gi} 
& $0.33 \pm 0.04$~\cite{Harada:1992np} \\
\hline 
\end{tabular} 
\caption{The predicted values of various QCD observables obtained from holographic calculations 
for $G=0.25$ (and $G=0$) with $f_\pi= 92.4$ MeV, $M_\rho=775.49$ MeV fixed, in comparison with the observed values.
} 
\label{fpi:mrho:ob:fixed:tab} 
\end{table}
We see that the agreement between holographic predictions and the observed values are much improved compared with the case $G=0$.
As was discussed in Ref.~\cite{Haba:2010hu}, nonzero gluonic effects with  $G\simeq 0.25$ 
are important to achieve such a simultaneous agreement of various QCD observables, most notably 
 $\langle \frac{\alpha_s}{\pi} G_{\mu\nu}^2 \rangle$ 
 and $M_{a_1}$.

Now we discuss the QCD gluonic effect on the hadronic leading order contribution to 
muon $g-2$, $a_\mu^{\rm HLO}$, which can be calculated through Eq.(\ref{amu:Hformula}). 
As was done for calculations of other observables, we take $\gamma_m=0$, 
$f_\pi=92.4$ MeV and $M_\rho=775.49$ MeV as inputs, then calculate the 
value of $a_\mu^{\rm HLO}$ as a function of $G$ to see gluonic effects on it.

\begin{figure}[h]
\begin{center}
 \includegraphics*[width=8.3cm]{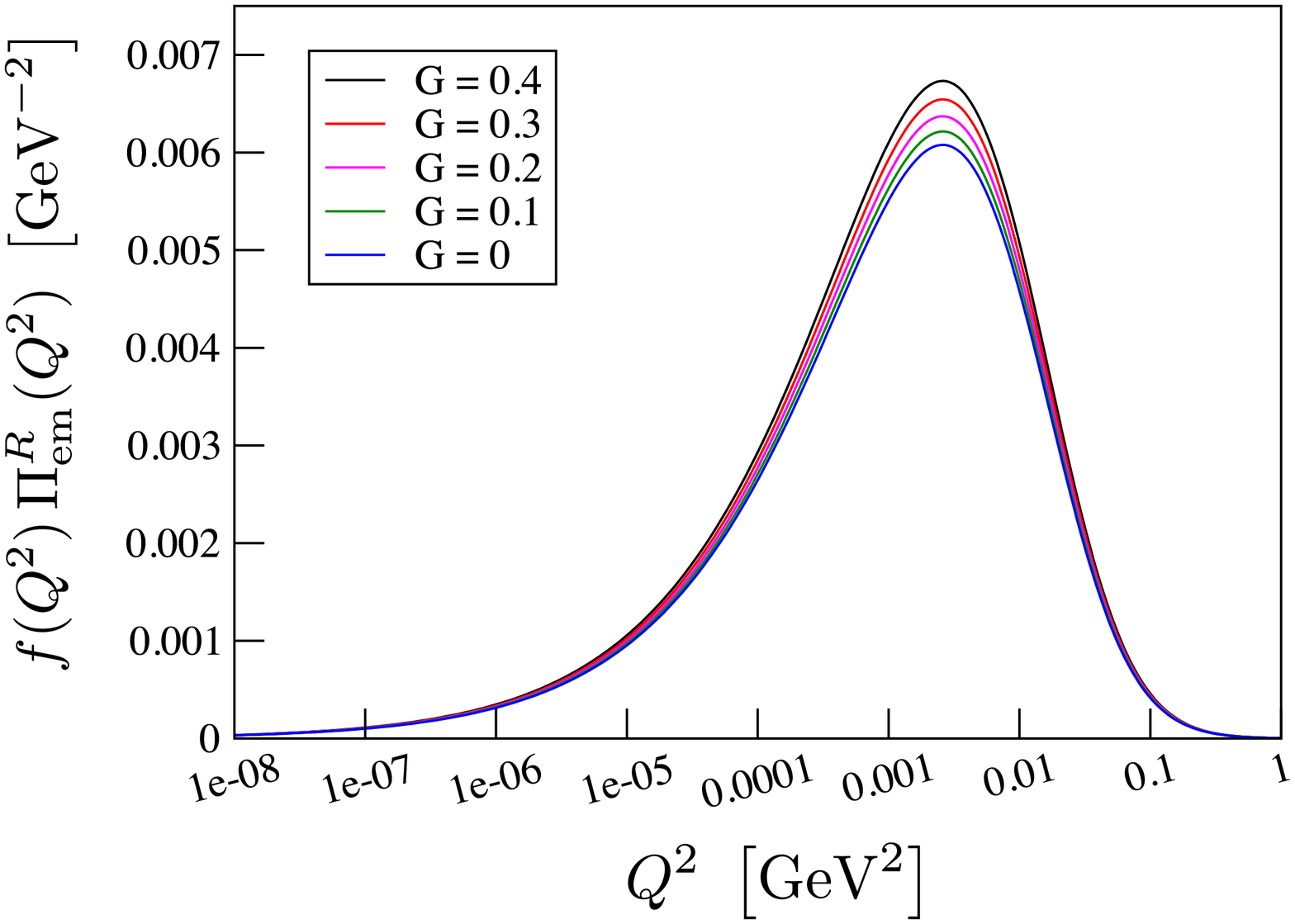}\ \ \ \ \  
  \includegraphics*[width=8.8cm]{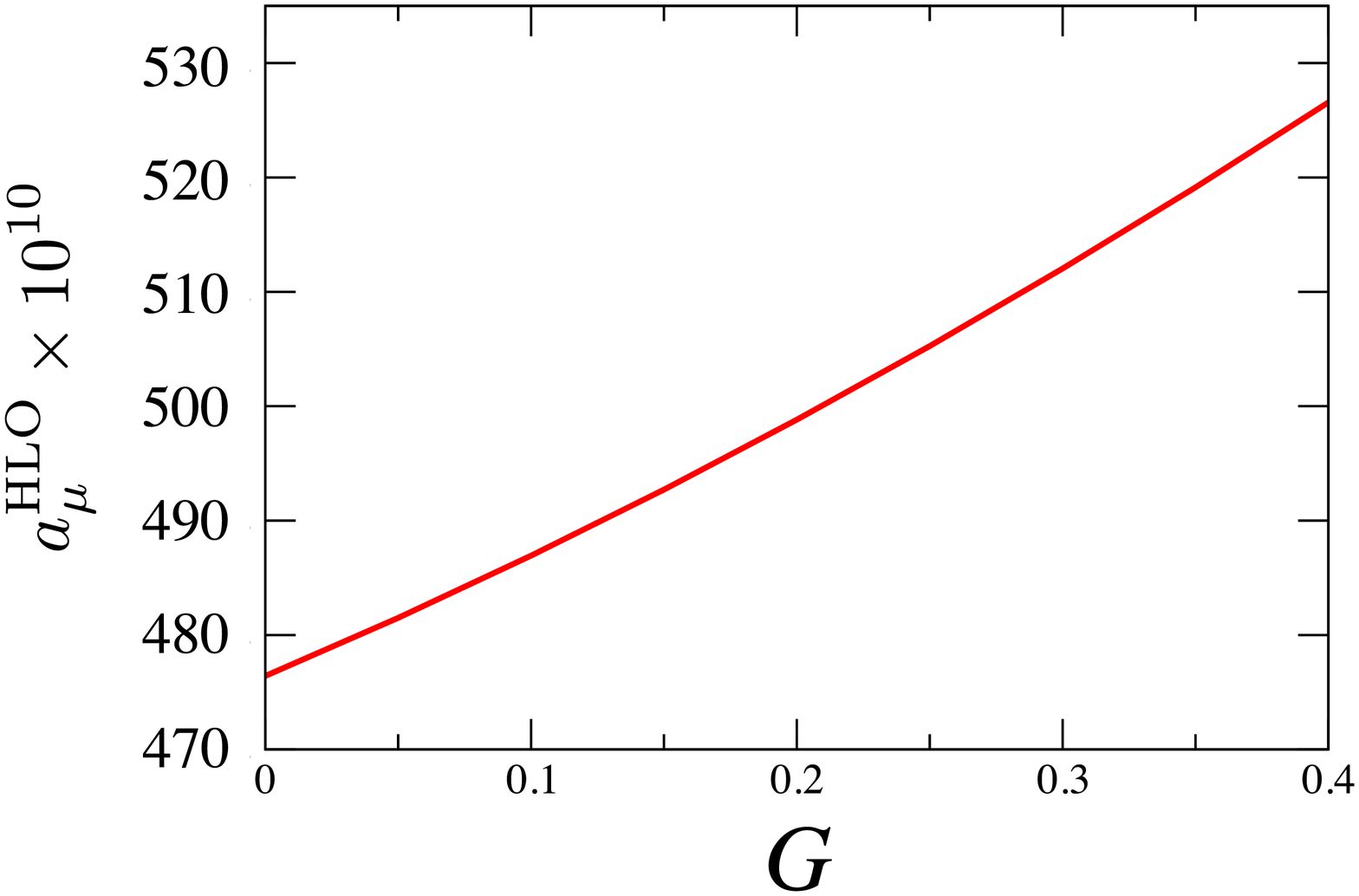}
\caption{ 
Left panel: The integrand in Eq.(\ref{amu:Hformula}) as a function of $Q^2$ varying 
$G$ from 0 to 0.4 with $f_\pi=92.4$ MeV and $M_\rho=775.49$ MeV fixed. 
Right panel:  
$a_\mu^{\rm HLO}$ as a function of $G$ for $N_f=2$ with 
$f_\pi=92.4$ MeV and $M_\rho=775.49$ MeV fixed. 
} 
\label{kernel}
\end{center} 
 \end{figure}

The left panel of Fig.~\ref{kernel} shows 
the integrand in Eq.(\ref{amu:Hformula}) as a function of $Q^2$ varying 
$G$ from 0 to 0.4 with $f_\pi=92.4$ MeV and $M_\rho=775.49$ MeV fixed. 
We see that as $G$ increases from 0 to 0.4, the peak value at around $Q^2 = m_\mu^2$ 
becomes larger, 
leading to the enhancement of $a_\mu^{\rm HLO}$. 
In the right panel of Fig.~\ref{kernel} we plot $a_\mu^{\rm HLO}$  as a function of $G$ 
for the case of $N_f=2$ with 
$f_\pi=92.4$ MeV and $M_\rho=775.49$ MeV fixed. 
As was expected from the enhancement of the integrand,  
the figure shows that the size of $a_\mu^{\rm HLO}$ becomes larger 
as $G$ increases. 
The values $a_\mu^{\rm HLO}$ at the optimal point ($G=0.25$) is 
estimated (in the chiral limit $m_u=m_d=m_s=0$) as 
\begin{eqnarray} 
a_\mu^{\rm HLO}|_{N_f=2} &\simeq & 
505 \times 10^{-10}
\qquad 
{\rm at} 
\qquad 
G = 0.25 
\,, \nonumber \\ 
a_\mu^{\rm HLO}|_{N_f=3} &\simeq & 
606 \times 10^{-10}
\qquad 
{\rm at} 
\qquad 
G = 0.25 
\,. \label{QCD:amu:predicted}
\end{eqnarray} 
The predicted value for $N_f=2$ above is 
in excellent agreement with a partial hadronic contribution to 
$a_\mu^{\rm HLO}$ estimated only from $\sigma(e^+e^- \to \pi^+\pi^-)$, 
$a_\mu^{\rm HLO}|_{\pi^+\pi^-}=(504.2 \pm 3.0) \times 10^{-10}$~\cite{Hagiwara:2011af}. 
 The size of $a_\mu^{\rm HLO}|_{N_f=3}$ 
is compared with the full hadronic contributions, $a_\mu^{\rm HLO}|_{\rm full}=(694.9 \pm 4.3)\times 10^{-10}$~\cite{Hagiwara:2011af}.  Agreements are quite impressive,
considering that our estimate is only at chiral limit $m_u=m_d=m_s=0$.

The above results can be compared to the value of $a_\mu^{\rm HLO}$ 
obtained without gluon condensation effect~\cite{Hong:2009jv}, which can be realized 
by setting $G=0$:
\begin{eqnarray} 
a_\mu^{\rm HLO}|_{N_f=2} &\simeq& 
 476 \times 10^{-10} 
\qquad 
{\rm at} 
\qquad 
G = 0 ,
\nonumber \\ 
a_\mu^{\rm HLO}|_{N_f=3} &\simeq& 
 571 \times 10^{-10} 
\qquad 
{\rm at} 
\qquad 
G = 0 
\,.  
\end{eqnarray}   
We see that the inclusion of gluonic effect results in enhancement of the value 
of $a_\mu^{\rm HLO}$ by about 6\%, and in both $N_f=2$ and $3$ cases, 
the agreement between holographic prediction and 
experimental value becomes better at $G=0.25$ compared to the results obtained 
from holographic model without gluonic effect.

Before closing this section, we show the results of similar holographic calculations for the 
HLO contribution to the anomalous magnetic moment of the electron and the tau lepton.
The formulae for $a_e^{\rm HLO}$ and $a_\tau^{\rm HLO}$  can be obtained 
simply by replacing $m_\mu$ in Eq.~(\ref{eq:weight}) by $m_e$ and $m_\tau$, respectively.
In Table~\ref{a-tau-mu-e:qcd}, we show the results in the case of $N_f=2, 3$ and 
$G=0, 0.25$. The values for $a_\mu^{\rm HLO}$ are also listed as well.
\begin{table}[h]
\begin{center}
\begin{tabular}{|l|c|c|c|}
\hline
& $a_e^{\rm HLO} \times 10^{14}$ & $a_\mu^{\rm HLO} \times 10^{10}$ & $a_\tau^{\rm HLO} \times 10^{8}$\\
\hline
$N_f$=2, G=0   & 125 & 476 & 230 \\
\hline
$N_f$=2, G=0.25  & 133 & 505 & 239 \\
\hline
$N_f$=3, G=0     & 150 & 571 & 276 \\
\hline
$N_f$=3, G=0.25  & 160 & 606 & 287 \\
\hline
\end{tabular}
\end{center}
\caption{Summary of holographic calculations for the HLO contribution to 
the anomalous magnetic moment of leptons. $f_\pi$=92.4 MeV and 
$M_\rho$=775.4 MeV are used as inputs. } 
\label{a-tau-mu-e:qcd}
\end{table}
The resultant value of $a_e^{\rm HLO}$ can be compared to the value 
$\sim186.6(\pm 1.1)\times10^{-14}$, which was obtained in Ref.~\cite{Nomura:2012sb} 
by using the same $e^+e^- \rightarrow$ hadrons data as those used in 
Ref.~\cite{Hagiwara:2011af}. As for $a_\tau^{\rm HLO}$, 
it is roughly $(200-400) \times 10^{-8}$. (See, 
for example, Table 1 in Ref.~\cite{Eidelman:2007sb} for a nice summary of various 
estimations of $a_\tau^{\rm HLO}$.)
It is remarkable that the holographic predictions of the HLO contribution to $g-2$ are 
quite consistent with the known values for all leptons. Considering the fact that the 
energy scales which are important for determining $a_e^{\rm HLO}$, 
$a_\mu^{\rm HLO}$ and $a_\tau^{\rm HLO}$ are $Q^2\sim m_e^2, m_\mu^2$ 
and $m_\tau^2$ respectively, the above mentioned good agreement 
indicates that the holographic calculation of $\Pi_{\rm em}(Q^2)$ is quite reliable in a 
 {\it wide range of 
the continuous space-like momentum}, not just  in the range of the (discrete) time-like momentum where the resonance parameters are fitted in the conventional holographic
studies in the zero-width approximation (large $N_c$ limit).

\section{Gluonic effects in walking technicolor} 
\label{sec:WTC}

If the EW symmetry is dynamically broken by WTC, it is natural to expect 
that there are techni-hadronic contributions to the anomalous magnetic moment of leptons 
in a way analogous to the QCD HLO contributions. In this section, 
we estimate such effects in WTC. 
Throughout calculations in this section, we take $\gamma_m=1$ for the bulk scalar mass term, 
instead of $\gamma_m=0$ for QCD in the previous section, 
so that the model reproduces desired walking behavior of WTC. To be concrete, 
we take the one-family model\cite{Farhi:1979zx} for the WTC as an example, 
in which case the number of techni-fermion is $N_f=8$, having $N_D=4$ weak doublet 
in the model. Actually, recent lattice results for $SU(3)$ gauge theories suggest that $N_f=8$ is a walking
theory~\cite{Aoki:2013xza}.
The techni-pion decay constant $F_\pi$ is related to the EW  
scale $v_{\rm EW}$ as 
$F_\pi = v_{\rm EW}/\sqrt{N_D}$, therefore, in the case of one-family WTC model, 
it is fixed to be $F_\pi = 246 {\rm GeV}/\sqrt{4} = 123$ GeV.
In order to see the dependence of physical observables on $G$ in WTC, we further fix
the holographic infrared scale $z_m^{-1}$ at typical values $z_m^{-1}=2,\,4,\,6 \,{\rm TeV}$, which roughly covers the phenomenologically interesting region of the $S$ parameter~\cite{Peskin:1990zt}, $0.1<S< 1.0$.
(see discussions below).

\begin{figure}[h]
\begin{center}
  \includegraphics[width=5.7cm]{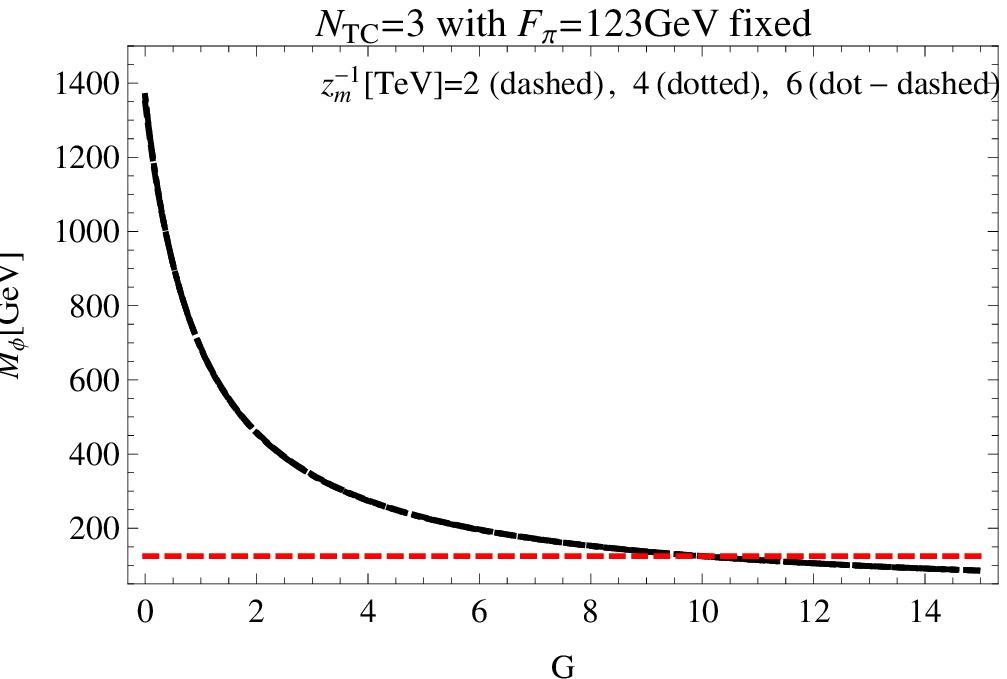}\
 \includegraphics[width=5.7cm]{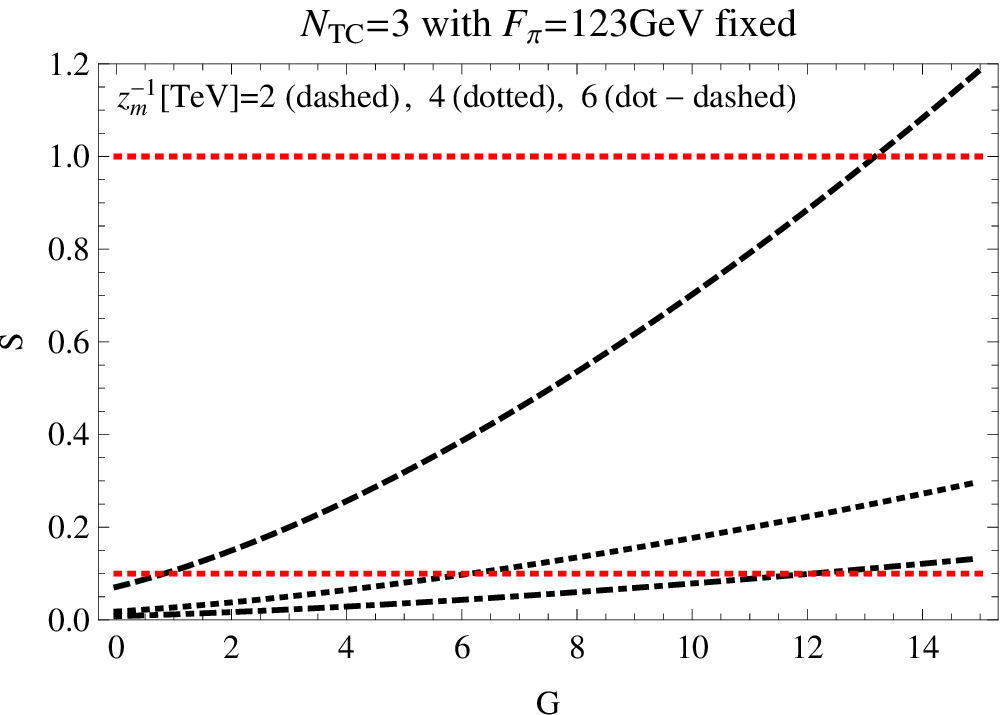}\
  \includegraphics[width=5.7cm]{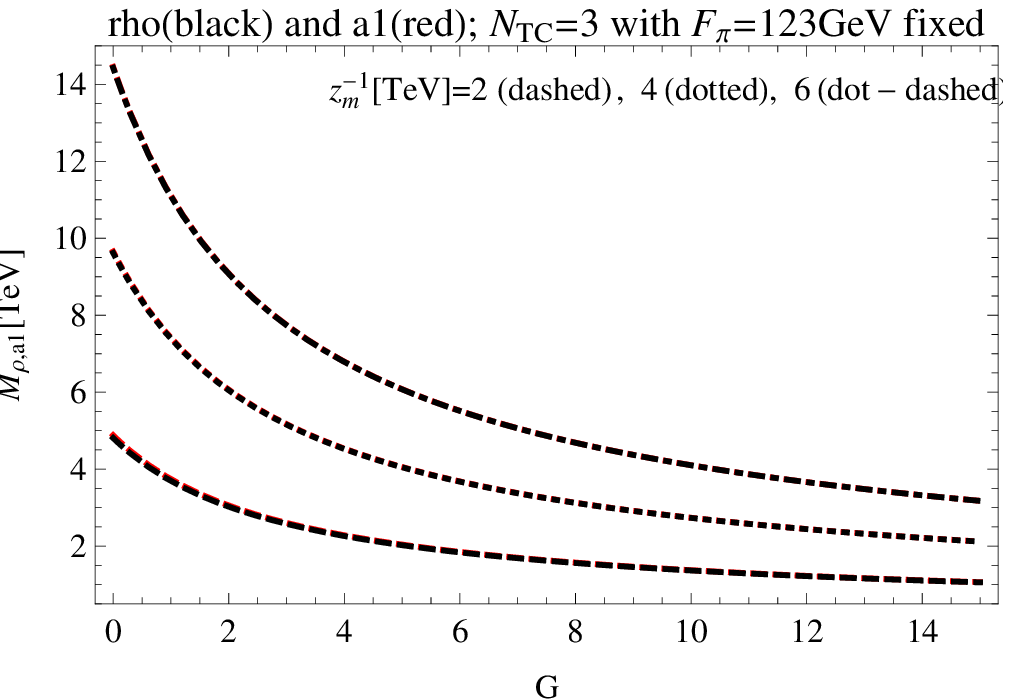}\vspace{4mm}\\
         \includegraphics[width=5.7cm]{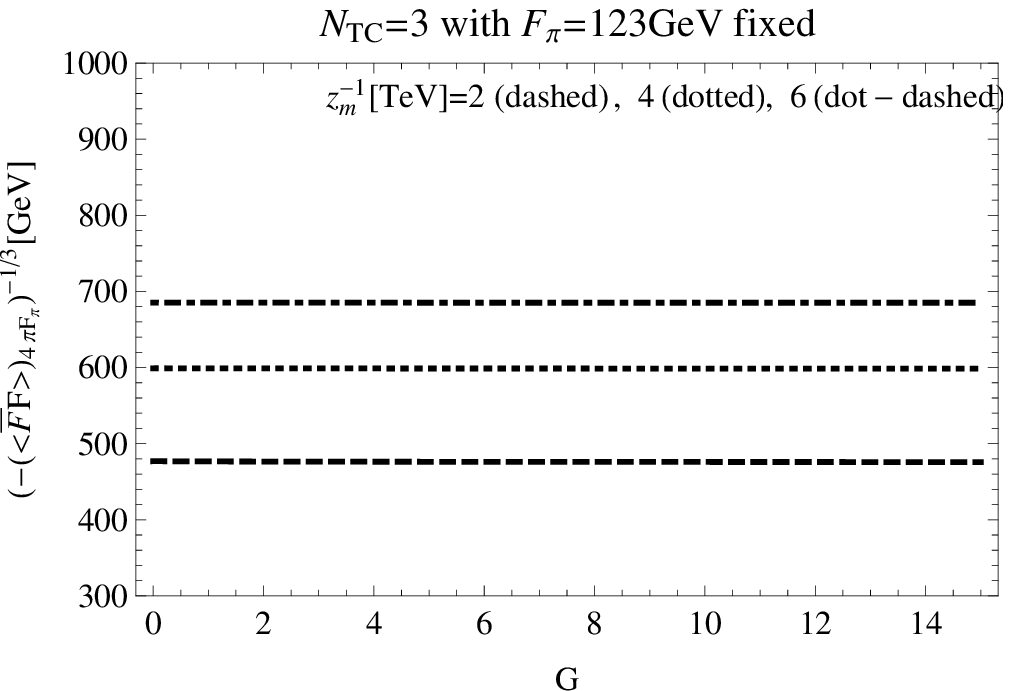}\ \ \
         \includegraphics[width=5.7cm]{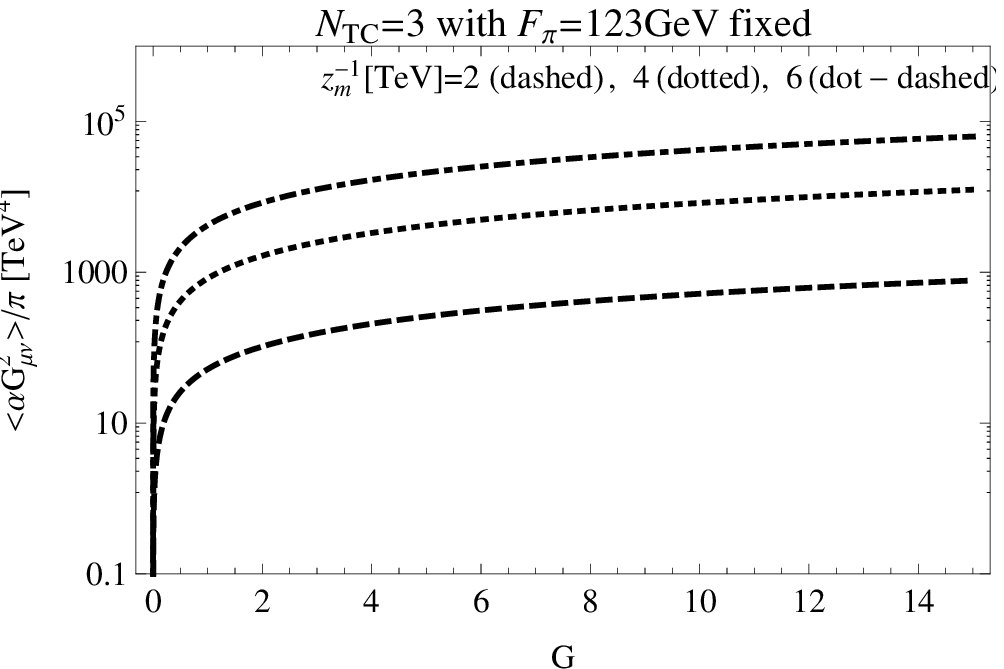}\ \ \
                   \includegraphics[width=5.7cm]{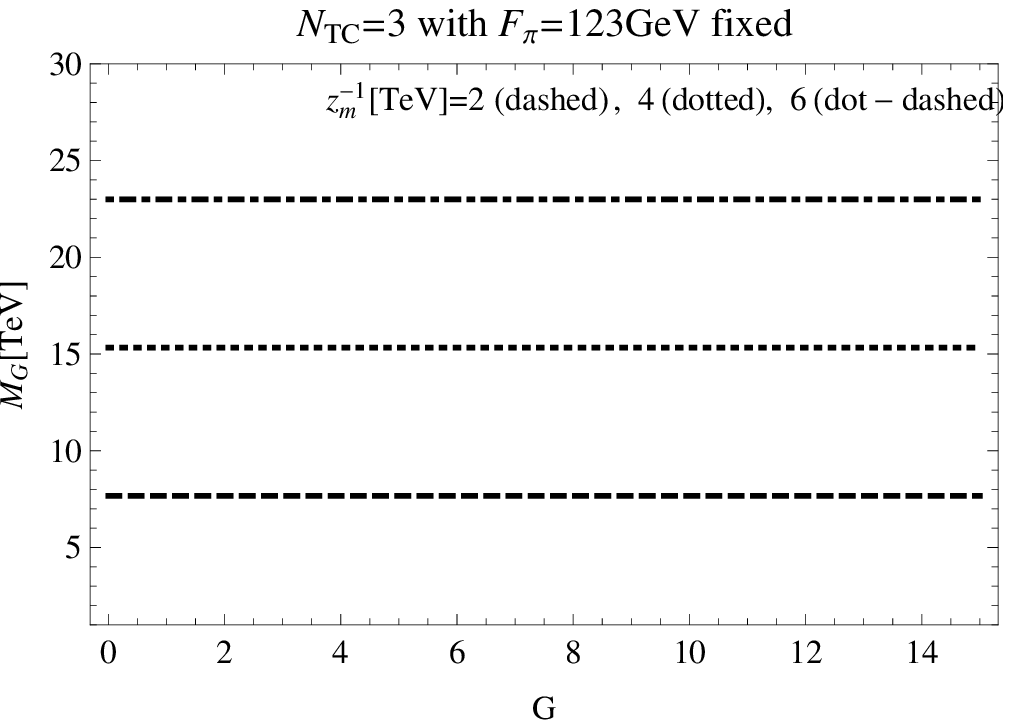} 
\caption{ 
Various quantities as functions of $G$ for the one-family WTC with 
$N_{\rm TC}=$3 and $F_\pi=123$ GeV fixed. Here, we have taken 
$z_m^{-1} = 2$ (dashed curve), $4$ (dotted curve) and $6$ (dot-dashed curve) TeV. 
Top-left: $M_\phi$  (Since three curves lie on top of one another, they are almost indistinguishable in the figure. 
The red dotted line indicates $M_\phi=$125 GeV), 
Top-center: $S$ parameter (The red dotted line indicates $S=0.1$), 
Top-right: $M_{\rho}$ (black curves) and $M_{a_1}$ (red curves) (Note that $\rho$ and $a_1$ are almost degenerate),
Bottom-left:  $(- \langle \bar{F}F \rangle_{\mu
=4\pi F_\pi})^{1/3}$,
Bottom-center: 
$\langle \frac{\alpha}{\pi} G_{\mu\nu}^2  \rangle$, 
Bottom-right:  $M_{G}$.
\label{N3-WTC}} 
\end{center} 
 \end{figure}

In Fig.~\ref{N3-WTC} we show the $G$ dependences of various observables  in the case of $N_{\rm TC}=3$:
the TD mass $M_\phi$ (Top-left), the $S$ parameter (Top-center), 
the masses of techni-$\rho$  ($M_\rho$) and techni-$a_1$ ($M_{a_1}$), degenerate each other (Top-right),  
the chiral condensate $\langle \bar F F \rangle$ (Bottom-left), the gluon condensate $\langle \frac{\alpha}{\pi} G_{\mu\nu}^2 \rangle$ (Bottom-center), 
and the techni-glueball mass $M_G$ (Bottom-right).
Remarkably enough, 
the TD mass $M_\phi$ (top-right figure) 
  dramatically changes from the order of 
TeV down to $100$ GeV as $G$ varies from 0 to $G=O(10)$~\cite{Haba:2010hu,Matsuzaki:2012xx}. 
Actually, we can obtain a sensible vanishing TD mass limit, {\it quite independently of $z_m^{-1}$}~\cite{Matsuzaki:2012xx}:
\begin{equation}
\frac{M_\phi}{4\pi F_\pi}\simeq 
\sqrt{
\frac{3}{N_{\rm TC}}
}
\frac{\sqrt{3}/2}{1+G} \rightarrow 0 \,\,\,(G\rightarrow \infty)\, .
\end{equation}
Thus in order to have a naturally light TD in the WTC case, we need the role of the gluon condensate 
more eminent than in QCD.
Since we identify the TD, flavor-singlet scalar $\phi$, as the 125 GeV boson discovered at the LHC~\cite{Chatrchyan:2012ufa}~\footnote{
 As was shown in Refs.~\cite{Matsuzaki:2012gd,Matsuzaki:2012xx}, 
 the TD in the one-family WTC indeed has the LHC signal consistent with the currently reported experimental data, 
notably explains the diphoton excess~\cite{Chatrchyan:2012ufa}. 
}, we refer to the value of $G$ which reproduces $M_\phi\simeq125$ GeV as the ``physical point".
In the top-left panel of Fig.~\ref{N3-WTC}, we indicated $M_\phi=125$ GeV as red-dashed 
line, so that we can easily find the physical point. From the figure, 
in the case of $N_{\rm TC}=3$, one can see that the value of 
$G\simeq 10$ at the physical point is 
rather insensitive to the values of $z_m^{-1}$.

As to the $S$ parameter, it increases as $G$ grows. However, 
we can see from Fig.~\ref{N3-WTC} that we can {\it freely adjust a small $S$ parameter, say $S<0.1$}, 
by increasing $z_m^{-1}$ (or equivalently increasing $M_\rho$), 
{\it without affecting the LHC phenomenology} of the physical point $M_\phi=125 \,{\rm GeV}$ (and also couplings) 
which is quite independent of $z_m^{-1}$~\cite{Matsuzaki:2012xx}.
Here we are particularly interested in the region, $0.1<S<1.0$, corresponding to 
the lighter $M_{\rho/a_1}$ accessible at the future LHC. Note that $S \propto F_\pi^2/M_\rho^2$~\cite{Haba:2010hu,Matsuzaki:2012xx}. 
Although $S=0.1$ is a phenomenologically viable benchmark value, 
there is a possibility 
that even if the WTC dynamics itself produces a large value of $S$, contributions 
coming from other part of the model such as the ETC interactions could partially cancel it in a way similar to 
the concept of fermion-delocalization effect studied in Higgsless models~\cite{Cacciapaglia:2004rb}. 
(In this sense $S>1.0$ might also be viable and even more interesting in view of having much lighter $M_{\rho/a_1}$ below TeV to be tested at the LHC.) 

\begin{figure}[h]
\begin{center}
  \includegraphics[width=7.0cm]{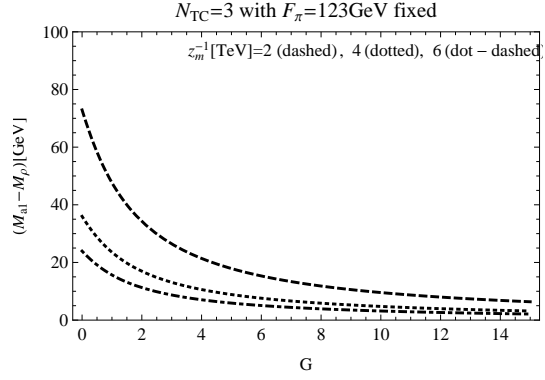}\
\caption{ $G$ dependence of the degeneracy of the mass of techni-$\rho$ and techni-$a_1$ for $N_{\rm TC}=3$ at $F_\pi=123$ GeV 
and $z_m^{-1}= 2,\ 4,\ 6$ TeV. 
\label{rhoa1}} 
\end{center} 
 \end{figure}

In accord with $S \propto F_\pi^2/M_\rho^2$, 
we see in  the Top-right figure in Fig.~\ref{N3-WTC} 
that $M_{\rho}$ and $M_{a_1}$  substantially decrease as the
value of $G$ increases, which would imply that the masses $M_{\rho}$ and $M_{a_1}$ at $G=0$ solely generated by the chiral condensate are drastically reduced by the large 
gluonic contributions. Also note the almost degenerate masses for all $G$ getting 
eventually degenerate at large $G$~\cite{Haba:2010hu,Matsuzaki:2012xx}. See Fig.\ref{rhoa1}.
It is also noted that this degeneracy does not imply 
a big cancellation of both contributions so as to yield a small $S$ parameter.

We can also do  similar calculations for the case of $N_{\rm TC}=4$ and $5$. 
The holographic parameters $(G, z_m^{-1},\,\xi)$ at the physical points $M_\phi=125 \,{\rm GeV}$ for $S=(0.1, \,0.3,\, 1.0)$  in the cases of  $N_{\rm TC}=3, 4$ and $5$  are summarized 
as~\cite{Matsuzaki:2012xx}: 
\begin{equation}
\begin{array}{cc}
G \simeq 10 ,\,\, z_m^{-1}[{\rm TeV}] = (5.3, \,3.1, \,1.7) ,\,\, \xi = (0.01,\, 0.02, \,0.04)& \qquad {\rm for} \qquad 
 N_{\rm TC}=3, \\ 
G \simeq 8.5,\,\, z_m^{-1}[{\rm TeV}] = (4.8, \,2.8,\, 1.5) ,\,\, \xi = (0.01,\, 0.03,\, 0.05) & \qquad {\rm for} \qquad N_{\rm TC}=4, \\ 
G \simeq 7.5,\,\, z_m^{-1}[{\rm TeV}] = (4.5,\, 2.6,\, 1.4) ,\,\, \xi = (0.02,\, 0.03,\, 0.05)   & \qquad {\rm for} \qquad N_{\rm TC}=5, \\ 
\end{array} 
\label{G:WTC}
\end{equation}
which is compared with Eq.(\ref{G:best}) in QCD.
In Table~\ref{WTC:tab} we make a list of the predicted values of various quantities at 
the physical point for the cases of $N_{\rm TC}=3, 4$ and 5 for the phenomenologically interesting values $S=(0.1, \,0.3,\, 1.0)$. 
\begin{table}[h] 
\begin{tabular}{|c||c|c|c|c|c|c|c|}
\hline 
 $N_{\rm TC}$ 
& $G$  
& $M_\phi$ [GeV] 
&  $M_{\rho}$ [TeV] 
 &  $M_{a_1}$ [TeV]
&  $M_{G}$ [TeV] 
&  $(- \langle \bar{F}F \rangle_{\mu
=4\pi F_\pi})^{1/3}$ [GeV] 
&  $\frac{1}{\pi} \langle \alpha G_{\mu\nu}^2 \rangle$ $[{\rm TeV}^4]$ \\ 
\hline \hline 
3 
& 10 & (124, 125, 125) & (3.6, 2.1, 1.1) & (3.6, 2.1, 1.2) & (20, 11, 5.4)  
& (658, 530, 423) & (2.6, 0.16, 0.0094) $\times 10^4$   \\
\hline 
4 
& 8.5 & (125, 125, 126)  & (3.6, 2.1, 1.1) & (3.6, 2.1, 1.2) & (20, 11, 5.4)  
& (628, 505, 405)& (2.6, 0.16, 0.0095) $\times 10^4$   \\ 
\hline 
5 
& 7.5 & (125, 125, 126)  & (3.6, 2.1, 1.1)  & (3.6, 2.1, 1.2) & (20, 11, 5.4)  
&(604, 486, 388) & (2.6, 0.16, 0.0095) $\times 10^4$   \\  
\hline 
\end{tabular} 
\caption{The predicted values of various observables in WTC with $N_{\rm TC}=3,4,5$ 
at the physical point shown in Eq.(\ref{G:WTC}) and $F_\pi= 123$ GeV.  
Three values 
in each parenthesis in the table correspond to the cases of $S=(0.1,\, 0.3,\, 1.0)$ 
from left to right, respectively.}  
\label{WTC:tab} 
\end{table}  

It should be noted that the value of $G$ at the physical point is quite 
large in the case of WTC compared to the case of QCD, whose 
physical point is $G=0.25$ (see Table~\ref{fpi:mrho:ob:fixed:tab} 
and discussion around Eq.~(\ref{G:best})). 
Thus the inclusion of the gluonic effects plays a vital role
for the holographic calculations of physical quantities in the case of WTC.
This is also the case for  the techni-HLO contributions 
to the anomalous magnetic moment of leptons as we will see below.

\begin{figure}[h]
\begin{center}
 \includegraphics*[scale=0.4]{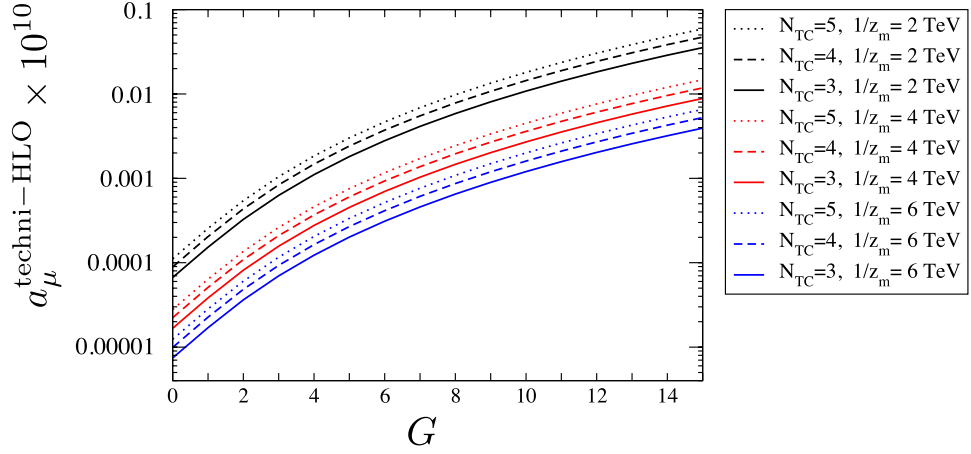}
\caption{ 
$a_\mu^{\rm techni-HLO}$ as a function of $G$ in the case of the one-family WTC for  
$N_{\rm TC}=$3(solid), 4(dashed) and 5(dotted) with $F_\pi=123$ GeV, $M_\phi=125$ GeV and $z_m^{-1}=$ 2 TeV (black), 4 TeV (red), and 6 TeV (blue)  fixed. 
\label{aHLO_TC}} 
\end{center} 
 \end{figure}

Let us now evaluate the gluonic effect on 
the techni-HLO contributions to the muon $g-2$. 
In Fig.~\ref{aHLO_TC}, we plot the results of holographic calculations of 
the techni-HLO contribution to the anomalous magnetic moment of the muon, 
$a_\mu^{\rm techni-HLO}$, as a function of $G$ in the case of the one-family WTC for  
$N_{\rm TC}=$3 (solid curve), 4 (dashed curve) and 5 (dotted curve) with 
$F_\pi=123$ GeV, $M_\phi=125$ GeV and $z_m^{-1}=$2 (black), 4 (red), 6 (blue) TeV. 
From this figure, we see a general trend that the value of 
$a_\mu^{\rm techni-HLO}$ monotonically increases as 
a function of $G$ for all the combinations of $N_{\rm TC}$ 
and $z_m^{-1}$: 
The physical point $G\simeq 10$ ($M_\phi=125\, {\rm GeV}$) implies about $10^2$ times enhancement compared with the $G=0$ value.

In Table~\ref{tab:atHLO-comp}, 
we list explicit values of $a_\mu^{\rm techni-HLO}$ 
at $G=0$ and those at the physical point for each combination of 
$N_{\rm TC}$ under the phenomenological constraint on the $S$ parameter (see Fig. \ref{N3-WTC} (Top-center panel) for G-dependence of $S$),  
so that we can easily see how much the 
value of $a_\mu^{\rm techni-HLO}$ at the physical point is enhanced by the gluon-condensation effect under such a constraint.
\begin{table}[h]
\begin{center}
\begin{tabular}{|l|c|}
\hline
\ \ \ \ \ $N_{\rm TC}=3$&$a_\mu^{\rm techni-HLO} \times 10^{10}$\\
\hline
\hline
$S=0.1,\ G=\ \, 0$ & 0.0000946 \\
\hline
$S=0.1,\ G=10$ & 0.00153 \\
\hline
\hline
$S=0.3,\ G=\ \, 0$ & 0.000298 \\
\hline 
$S=0.3,\ G=10$ & 0.00460 \\
\hline
\hline
$S=1.0,\ G=\ \, 0$ & 0.00126 \\
\hline
$S=1.0,\ G=10$ & 0.0155 \\
\hline
\end{tabular}
\ \ \ \ \ \ 
\begin{tabular}{|l|c|}
\hline
\ \ \ \ \ $N_{\rm TC}=4$&$a_\mu^{\rm techni-HLO} \times 10^{10}$\\
\hline
\hline
$S=0.1,\ G=\ \ 0$ &0.000125  \\
\hline
$S=0.1,\ G=8.5$ & 0.00159 \\
\hline
\hline
$S=0.3,\ G=\ \ 0$ & 0.000390 \\
\hline
$S=0.3,\ G=8.5$ & 0.00477 \\
\hline
\hline
$S=1.0,\ G=\ \ 0$ & 0.00152 \\
\hline
$S=1.0,\ G=8.5$ & 0.0160 \\
\hline
\end{tabular}
\ \ \ \ \ \ 
\begin{tabular}{|l|c|}
\hline
\ \ \ \ \ $N_{\rm TC}=5$&$a_\mu^{\rm techni-HLO} \times 10^{10}$\\
\hline
\hline
$S=0.1,\ G=\ \ 0$ & 0.000156 \\
\hline
$S=0.1,\ G=7.5$ & 0.00164 \\
\hline
\hline
$S=0.3,\ G=\ \ 0$ & 0.000482 \\
\hline
$S=0.3,\ G=7.5$ & 0.00494 \\
\hline
\hline
$S=1.0,\ G=\ \ 0$ & 0.00180 \\
\hline
$S=1.0,\ G=7.5$ & 0.0166 \\
\hline
\end{tabular}
\begin{tabular}{|r|c|c|}
\end{tabular}
\begin{tabular}{|r|c|c|}
\end{tabular}
\end{center}
\label{tab:atHLO-comp}
\caption{Comparison of values of $a_\mu^{\rm techni-HLO}$ of the 
one-family WTC at 
$G=0$ and at the physical point for each combination of $N_{\rm TC}$ 
and $S$. Here,  $F_\pi$ and $M_\phi$ are fixed to be
$F_\pi=123$ GeV and $M_\phi=125$ GeV, respectively.
}
\end{table}%
From this table, we see that, in most cases, the value of 
$a_\mu^{\rm techni-HLO}$ at the physical point is more than 10 times 
larger than the value at $G=0$ for $S=0.1$. Also, for any given combination of $N_{\rm TC}$ 
and $G$, the value of $a_\mu^{\rm techni-HLO}$ is  
enhanced by another factor of more than 10 when we change the constraint on $S$ 
from $0.1$ to $1.0$. (If we are allowed to take even larger $S$ such as $S>1.0$ 
for the reason we mentioned before, we would get further drastically large
enhancement.) 

\begin{figure}[h]
\begin{center}
 \includegraphics*[scale=0.34]{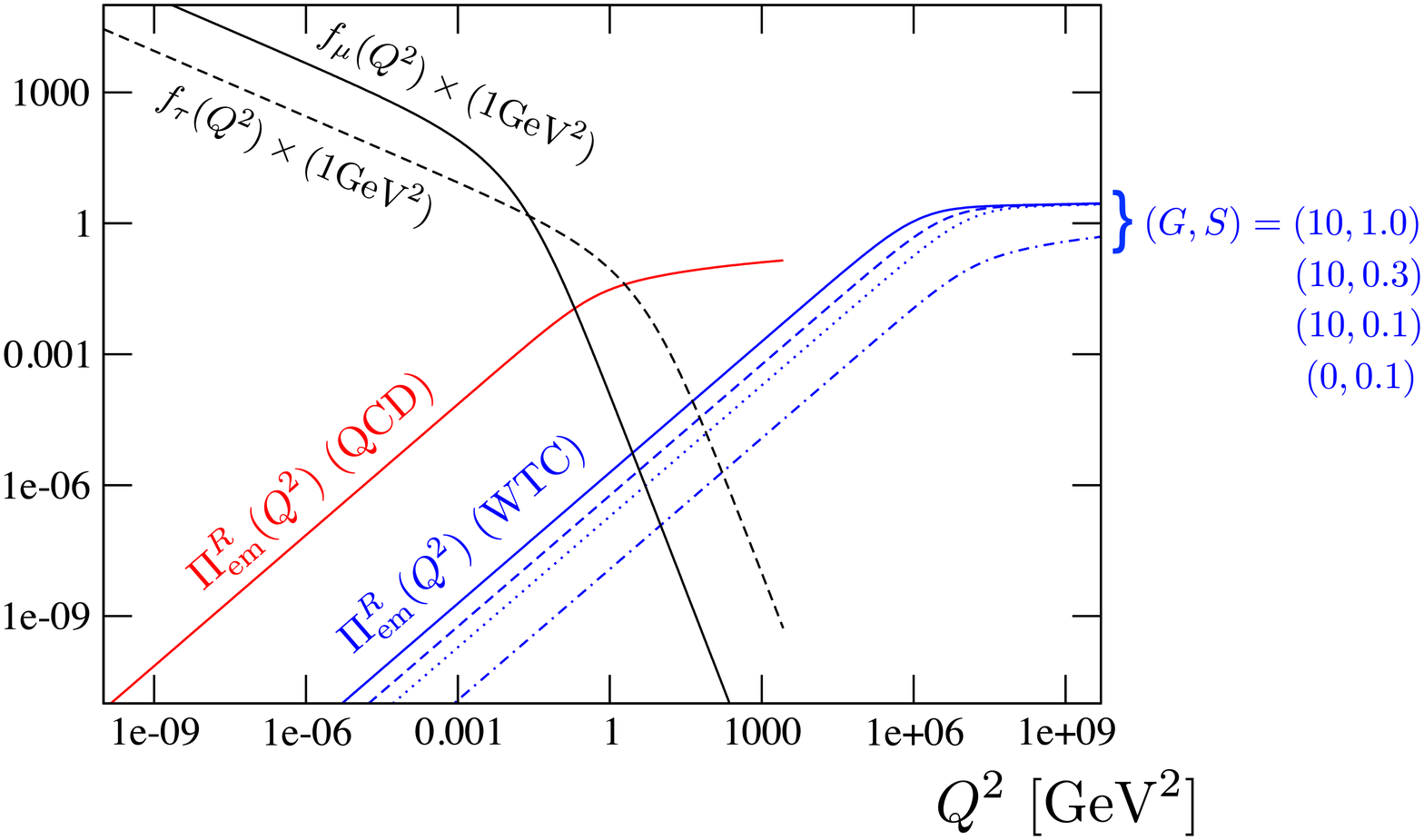}
\caption{Black solid (dashed) curve represents weight function $f(Q^2)$ in 
Eq.~(\ref{eq:weight}) for the muon (tau lepton) in unit of ${\rm GeV}^{-2}$.
Red solid curve represents renormalized electromagnetic current correlator 
$\Pi^R_{\rm em}(Q^2)$ which comes from vacuum polarization due to QCD. 
Blue curves are contributions to $\Pi^R_{\rm em}(Q^2)$ from WTC dynamics 
with $N_{TC}=3$ and $F_\pi=123$ GeV fixed: 
Dotted and dashed-dotted curves correspond to the cases of   
$(G, S)
= (10, 0.1)$ and $(0, 0.1)$, respectively. This enhancement due to $G =0 \rightarrow 10$ is more eminent as illustrated by the solid and dashed-dotted curves, $(G, S)=(10, 1.0), (10, 0.3)$, respectively.}
\label{fig:Pi_WTC} 
\end{center} 
 \end{figure}

These trends are visibly understood from 
Fig.~\ref{fig:Pi_WTC}, where we plotted contributions to $\Pi_{\rm em}^R(Q^2)$ 
from QCD, as well as that from WTC dynamics in comparison between 
$(G, S) = 
(10, 0.1)$ and $(0, 0.1)$ to show the enhancement of $a_\mu^{\rm techni-HLO}$
simply due to the increased gluonic effects  $G$ for the same S value. We also included the data for $(G, S) = (10, 1.0), (10, 0.3)$ just for 
illustration of  this
enhancement by the change from $G=0$ to $G=10$  
could be amplified if we relax the S parameter constraint.  
In the figure, we also plotted 
weight function in Eq.~(\ref{eq:weight}) for the muon (denoted as $f_\mu(Q^2)$ 
in the figure) in unit of ${\rm GeV}^{-2}$. The energy scale at which the slope 
(in the log-log plot) of the weight function changes roughly corresponds to the scale 
where the integrand in Eq.~(\ref{eq:weight-1}) has its peak, namely $Q^2\sim m_\mu^2 \simeq 0.01\, {\rm GeV}^2$. 
Therefore, very roughly 
speaking, the contribution to $a_\mu^{\rm (techni-)HLO}$ is proportional to the 
magnitude of $\Pi_{\rm em}^R(Q^2)$ at this scale.
It should be noted that 
$\Pi_{\rm em}^R(Q^2)|_{\rm QCD}/\Pi_{\rm em}^R(Q^2)|_{\rm WTC}$ is almost constant 
at around $Q^2\sim m_\mu^2 \simeq 0.01\, {\rm GeV}^2$, 
while the ratio becomes larger at $Q^2\sim m_\tau^2 \simeq 3\,{\rm GeV}^2$. 
This implies that the WTC contribution gets relatively more important in the case of tau $g-2$.  
In Table~\ref{tab:tau}, 
we summarize the size of contributions to $a_\mu^{\rm (techni-)HLO}$ and 
$a_\tau^{\rm (techni-)HLO}$ from QCD and WTC at the physical point. 
As for WTC contributions, two cases (i.e., $S=0.1$ and $1.0$) are shown in the table.
\begin{table}[h]
\begin{center}
\begin{tabular}{|l|c|c|}
\hline
& $a_\mu^{\rm (techni-)HLO} \times 10^{10}$ & $a_\tau^{\rm (techni-)HLO} \times 10^{8}$\\
\hline
\  QCD ($N_f=3$)\hspace{2.4mm} : $f_\pi=92.4\,{\rm MeV},\, M_\rho=775.49 \,{\rm MeV},\, G=0.25$\ \ 
&606 & 287 \\
\hline
\,WTC ($N_{TC}=3$) : $F_\pi=123\,{\rm GeV},\ S=0.1,\ G=10$& 
 0.00153 & 0.00433\\
\hline
\,WTC ($N_{TC}=3$) : $F_\pi=123\,{\rm GeV},\ S=1.0,\ G=10$&
 0.0155 & 0.0437\\
\hline
\end{tabular}
\end{center}
\caption{Summary of holographic calculations of 
contributions to $a_\mu^{\rm (techni-)HLO}$ and 
$a_\tau^{\rm (techni-)HLO}$ from QCD and WTC dynamics at the physical point. 
Two cases, $S=0.1$ and $1.0$, are shown for WTC contribution.}
\label{tab:tau}
\end{table}
As we expect, 
$a_\tau^{\rm HLO}/a_\tau^{\rm techni-HLO}$ is  about 6 times larger 
compared to the ratio in the case of muon $g-2$.


\section{Discussion and Conclusions}
\label{sec:conclusions}

We have shown the results of holographic 
calculations of the (techni-)HLO contributions to the anomalous magnetic moment $g-2$
of leptons. It was shown that, in the case of the QCD contribution, it was enhanced 
by about 6\% due to the proper inclusion of the gluon-condensation effect ($G=0.25$) compared 
to the estimate ignoring it ($G=0$)~\cite{Hong:2009jv}, leading to a better agreement with the known value determined by the experimental data~\cite{Hagiwara:2011af}: 
\begin{eqnarray} 
a_\mu^{\rm HLO}|_{N_f=2} &\simeq & 
505 \times 10^{-10}
\quad  
(a_\mu^{\rm HLO}|_{\pi^+\pi^-}=(504.2 \pm 3.0) \times 10^{-10})
\,, \\
\nonumber \\
a_\mu^{\rm HLO}|_{N_f=3} 
&\simeq&
606 \times 10^{-10}\, \quad
(a_\mu^{\rm HLO}|_{\rm full}=(694.9 \pm 4.3)\times 10^{-10})
\,. 
\end{eqnarray} 
Considering that our estimate was at the chiral limit $(m_u=m_d=m_s=0)$, the agreement is rather impressive.

In the case of WTC, where the gluonic effect plays more significant role compared to the case of QCD 
to reproduce the observed 125 GeV scalar boson, the gluon-condensation effects 
are more dramatic for the muon $g-2$:
The value of the techni-HLO contribution at physical value $G\simeq 10$ is more than 100 times larger compared to the estimate ignoring the gluonic effect($G=0$), if we allow rather large value of $S$ up to $S<1.0$.

It is quite interesting that the contributions to 
the anomalous magnetic moment from the vacuum polarization of 
the electromagnetic current is several orders of magnitude larger than the 
naive scale-up estimate, which is obtained by simply multiplying the ratio of 
the squares of typical scales of QCD ($f_\pi \simeq $ 92 MeV) to one-family model ($F_\pi \simeq$ 123 GeV) 
to the value of $a_\mu^{\rm HLO}$, yielding a value of $\simeq 3\times 10^{-14}$.  
This value roughly coincides with the value of holographic calculation with 
$G=0$, $S=0.3$ (see Table~\ref{tab:atHLO-comp}).

Though this kind of significant enhancement is quite interesting theoretically, 
the phenomenological interest is whether WTC contribution is visible or not. 
Even if there is an enhancement by several orders of magnitudes compared to 
the naive estimate, the contribution from the WTC dynamics to the $g-2$ is 
still negligibly small compared to the QCD contribution. The magnitude of 
WTC contribution to the muon $g-2$ is, at most, order of $10^{-12}$ for the cases 
of setup investigated in this paper. This value is still quite small compared to the 
current discrepancy between the experimental value 
of the muon $g-2$ and the standard model prediction of it:
$\delta a_\mu\equiv 
a_\mu^{\rm Exp}-a_\mu^{\rm SM}=(26.1\pm 8.0) \times 10^{-10}$~\cite{Hagiwara:2011af,Bennett:2006fi}. 
Therefore, 
it seems rather 
unlikely that the contributions from WTC 
dynamics can explain the current 3.3 $\sigma $ deviation between experiment and the SM prediction.

In conclusion we have shown, in a bottom-up holographic model for QCD/WTC, that the amount of gluonic effects 
to realize the physical point ($G=0.25$ for QCD and $G=10$ for WTC with $N_{\rm TC}=3$) makes significant effects enhancing the (techni-) HLO contributions to the lepton $g-2$.  Since the introduction of the gluonic effects in this model is just  to make the model consistent with the high energy region of the QCD, it is a highly nontrivial test of this holographic model whether or not the effects also improve the agreements with the low energy hadron physics, particularly in 
the most relevant momentum region $Q^2=-q^2 \sim m_l^2$ for $(l=e,\mu,\tau)$ (see Fig.\ref{fig:Pi_WTC}), which is far infrared region for QCD. Note that the
quantity studied in this paper is in the space-like momentum region $Q^2>0$ which is theoretically more tractable without limitation of the zero-width constraint of large
 $N_c$ limit where the holography is so far justified, in contrast to the resonance phenomenology in the time-like region.  Our work presents an explicit example of a holographic model which successfully 
 reproduces the QCD physics in all energy region from the very ultraviolet region for OPE all the way down to such a deep infrared region in the space-like momentum. The vital role of the gluon condensate in WTC in connection with the light scalar composite can be clarified by the future lattice studies such as an extension of the work which observed a light flavor-singlet scalar meson in large $N_f$ QCD~\cite{Aoki:2013zsa}. 

\section*{Acknowledgments}

This work was supported by 
the JSPS Grant-in-Aid for Scientific Research (S) \#22224003 and (C) \#23540300 (K.Y.).

\end{document}